\documentclass[aps,pre,superscriptaddress,twocolumn,longbibliography,groupedaddress]{revtex4-2}

\usepackage[colorlinks]{hyperref}
\hypersetup{
  linkcolor = {blue},
  citecolor = {blue},
  urlcolor = {cyan}
}

\usepackage{float}
\usepackage{graphicx}
\usepackage{caption}
\usepackage{subcaption}
\usepackage{amsmath}
\usepackage{amsthm,amssymb}
\usepackage{amsfonts}
\usepackage{dcolumn}
\usepackage{bm}
\usepackage{epstopdf}
\usepackage{algorithm}
\usepackage{algpseudocode}

\usepackage{epsfig}
\usepackage{mathrsfs}
\usepackage{multirow}
\usepackage{tabularx}
\usepackage{mathptmx}%
\usepackage[all]{xy}
\usepackage{pbox}
\usepackage{verbatim}
\usepackage{mathtools}
\usepackage{tikz,pgfplots}
\usepackage{xfrac}
\usepackage{cleveref}
\usepackage{booktabs}
\usepackage{physics}
\usepackage{siunitx}
\usepackage{url}
\usepackage{pict2e}
\usepackage{tikz,tikz-3dplot}
\usetikzlibrary{perspective}
\usepackage{xkeyval}
\usepackage{xcolor}
\usepackage[T1]{fontenc}
\usepackage{halloweenmath}
\usepackage{curve2e}
\usepackage{xtab,afterpage}
\usepackage{balance}
\usepackage{soul}
\usepackage[version=4]{mhchem}
\usepackage{braket}

\newcommand{\opcatMzzzz}[1]{{#1}_{\mathrm{zzzz}}}

\makeatletter

\@ifdefinable\SuCmathpictvertex{} 
\@ifdefinable\@SuC@reserved@dimen{\newdimen\@SuC@reserved@dimen}

\newenvironment*{@SuC@math@picture}[8]{%
  \def\SuCmathpictvertex{\circle*{#6}}%
  \setlength\unitlength{\fontdimen 22 #5\tw@}%
  \setlength\@SuC@reserved@dimen{#7\unitlength}%
  \kern\@SuC@reserved@dimen
  \@HwM@d@pict@strut{#2}%
  \picture(#3,#1)(#4,-1)%
    \roundcap
    \roundjoin
    \linethickness{#8\@HwM@thickness@units@for #5}%
}{%
  \endpicture
  \kern\@SuC@reserved@dimen
}
\newcommand*\@SuC@general@pict[9]{%
  \begin{@SuC@math@picture}%
            {#2}{#3}
            {#4}{#5}
            #6
            {#7}
            {#8}
            {#9}
    #1%
  \end{@SuC@math@picture}%
}
\newcommand*\@SuC@math@version@shunt[7]{%
  \@HwM@choose@thicknesses{\@SuC@general@pict {#1}{#2}{#3}{#4}{#5}#7}%
      {{.8}{.4}{}}
      {{1}{.5}{1.5}}
}

\newcommand*\DeclareNewSuCMathPict[6]{%
  \newcommand*{#1}{%
    \@HwM@general@ordinary@symbol
      {\@SuC@math@version@shunt {#6}{#2}{#3}{#4}{#5}}%
  }%
}

\makeatother

\DeclareNewSuCMathPict{\trigonleft}
            {3}{1}  
            {4}{-2} 
{
    \polygon(45:2)(135:2)(225:2)%
    \put (45:2){\SuCmathpictvertex}%
    \put(135:2){\SuCmathpictvertex}%
    \put(225:2){\SuCmathpictvertex}%
}

\DeclareNewSuCMathPict{\trigonright}
            {3}{1}  
            {4}{-2} 
{
    \polygon(45:2)(135:2)(315:2)%
    \put (45:2){\SuCmathpictvertex}%
    \put(135:2){\SuCmathpictvertex}%
    \put(315:2){\SuCmathpictvertex}%
}

\DeclareNewSuCMathPict{\trigonmiddle}
            {3}{1}  
            {4}{-2} 
{
    \polygon(45:2)(135:2)(270:2)%
    \put (45:2){\SuCmathpictvertex}%
    \put(135:2){\SuCmathpictvertex}%
    \put(270:2){\SuCmathpictvertex}%
}

\DeclareNewSuCMathPict{\tetrapleuro}
            {3}{1}  
            {4}{-2} 
{
    \polygon(45:2)(135:2)(270:2)%
    \put (45:2){\SuCmathpictvertex}%
    \put (90:1.4){\SuCmathpictvertex}%
    \put(135:2){\SuCmathpictvertex}%
    \put(270:2){\SuCmathpictvertex}%
}

\DeclareNewSuCMathPict{\pyramid}
            {3}{1}  
            {4}{-2} 
{
    \polygon(290:2)(40:2)(110:2)(10:2)%
}

\DeclareNewSuCMathPict{\Isingdiagonalsecond}
            {3}{1}  
            {4}{-2} 
{
    \polygon(45:2)(225:2)%
    \put (45:2){\SuCmathpictvertex}%
    \put(225:2){\SuCmathpictvertex}%
}

\DeclareNewSuCMathPict{\Isingdiagonal}
            {3}{1}  
            {4}{-2} 
{
    \polygon(135:2)(315:2)%
    \put (135:2){\SuCmathpictvertex}%
    \put(315:2){\SuCmathpictvertex}%
}

\DeclareNewSuCMathPict{\Isingvertical}
            {3}{1}  
            {4}{-2} 
{
    \polygon(135:2)(225:2)%
    \put (135:2){\SuCmathpictvertex}%
    \put(225:2){\SuCmathpictvertex}%
}

\newlength\figureheight 
\newlength\figurewidth 

\begin{document}

\title{Spin models from nonlinear cellular automata}

\author{Konstantinos Sfairopoulos}
\email{ksfairopoulos@gmail.com}
\affiliation{School of Physics and Astronomy, University of Nottingham, Nottingham, NG7 2RD, UK}
\affiliation{Centre for the Mathematics and Theoretical Physics of Quantum Non-Equilibrium Systems,
University of Nottingham, Nottingham, NG7 2RD, UK}
\author{Luke Causer}
\affiliation{School of Physics and Astronomy, University of Nottingham, Nottingham, NG7 2RD, UK}
\affiliation{Centre for the Mathematics and Theoretical Physics of Quantum Non-Equilibrium Systems,
University of Nottingham, Nottingham, NG7 2RD, UK}
\author{Jamie F. Mair}
\affiliation{School of Physics and Astronomy, University of Nottingham, Nottingham, NG7 2RD, UK}
\affiliation{Centre for the Mathematics and Theoretical Physics of Quantum Non-Equilibrium Systems,
University of Nottingham, Nottingham, NG7 2RD, UK}
\author{Stephen Powell}
\affiliation{School of Physics and Astronomy, University of Nottingham, Nottingham, NG7 2RD, UK}
\affiliation{Centre for the Mathematics and Theoretical Physics of Quantum Non-Equilibrium Systems,
University of Nottingham, Nottingham, NG7 2RD, UK}
\author{Juan P. Garrahan}
\affiliation{School of Physics and Astronomy, University of Nottingham, Nottingham, NG7 2RD, UK}
\affiliation{Centre for the Mathematics and Theoretical Physics of Quantum Non-Equilibrium Systems,
University of Nottingham, Nottingham, NG7 2RD, UK}

\begin{abstract}
We study classical and quantum spin models derived from one-dimensional cellular automata (CA) with nonlinear update rules, focusing on rules 30, 54 and 201. We argue that the classical models, defined such that their ground states correspond to allowed trajectories of the CA, are frustrated and can be described in terms of local defect variables.
Including quantum fluctuations through the addition of a transverse field, we study their ground state phase diagram and quantum phase transitions. We show that the nonlinearity of the CA rule leads to a quantum order-by-disorder mechanism, which selects a particular (rule-dependent) spatial structure for small transverse fields, with spontaneous breaking of the translation symmetry in some cases. Using numerical results for larger fields, we also observe a first-order quantum phase transition into a quantum paramagnet, as in previous studies of spin models based on linear CA rules. 
\end{abstract}

\maketitle

\section{Introduction}

Frustration refers to the phenomenon where the degrees of freedom of a system cannot simultaneously satisfy all of their interactions \cite{liebmann1986statistical,diep2020frustrated,2011introduction,2004quantum}. The origin of this effect can be twofold: geometrical \cite{moessner2006geometrical} or due to the interactions themselves \cite{diep2020frustrated}. A classic paradigm of geometric frustration consists of Ising spins on a triangular lattice interacting through antiferromagnetic, two-body terms. Minimization of the free energy of this classical system leads to an extensive ground state degeneracy, which was first studied by Wannier \cite{wannier1950antiferromagnetism.}. The restriction to Ising spins is in general unnecessary, since effects of frustration are equally encountered for Potts or vector spin systems \cite{yoshimori1959a-new-type,elliott1961phenomenological,kaplan1961some,villain1980order}.

Frustration often leads to an extensive (or subextensive) ground state degeneracy. Placing these systems under the effect of thermal or quantum fluctuations gives rise to a wealth of phenomena depending on whether the degeneracy is (fully or partially) lifted or if it persists \cite{diep2020frustrated}. In the former category, magnetization plateau structures might be encountered where the lifting of the degeneracy involves some kind of (spontaneous) symmetry breaking mechanism \cite{2011introduction}. At the same time, fluctuations do not always act destructively; the stabilisation of a part of the ground state degeneracy that possesses the softest fluctuations might lead to a ``fluctuation-induced'' ordering or order-by-disorder (ObD) \cite{villain1980order,henley1987ordering,henley1989ordering,chubukov1992order,chalker1992hidden}. This ordering mechanism can be equally of thermal (thObD) or quantum (qObD) origin. A subclass of this mechanism includes the possibility of the selection of a disordered or (cooperative) paramagnetic state in what is called ``disorder-by-disorder'' (DbD) \cite{moessner2001ising,moessner2001magnets,powalski2013disorder,narasimhan2024simulating}. On the other hand, a cooperative paramagnet or quantum spin liquid might be formed, where the spin degrees of freedom remain disordered but fluctuations are strongly correlated \cite{moessner1998low-temperature,moessner1998properties,savary2016quantum,zhou2017quantum,knolle2019a-field}. Note that the connectedness of the topology of the ground state manifold in this last case is of paramount importance \cite{moessner2000two-dimensional,moessner2001magnets} for the existence of the spin liquid phase.

A big part of the literature has focused on the study of the above phenomena in two paradigmatic models of condensed matter, the Heisenberg antiferromagnet \cite{moessner1998low-temperature,moessner1998properties,starykh2015unusual} and the transverse field Ising model \cite{moessner2000two-dimensional,moessner2001ising,moessner2001magnets}, placed on different frustrated lattices including the triangular, with short- \cite{moessner2001ising,dublenych2013ground} or long-range interactions \cite{koziol2024order-by-disorder}, or in buckled colloidal monolayers \cite{han2008geometric,shokef2009stripes,shokef2011order}, the kagome \cite{moessner2000two-dimensional,moessner2001ising,narasimhan2024simulating}, the ruby lattice\cite{duft2024order-by-disorder} or the bathroom tile \cite{hearth2022quantum}. Recently, these effects have been proposed to occur in systems of Rydberg atoms \cite{wu2024programmable}, although the quest for their experimental observation is much older \cite{gardner2010magnetic,struck2011quantum}. Ref.~\cite{gardner2010magnetic} analyzed a variety of magnetic pyrochlore oxides and the effects of frustration on them, while also reviewing another mechanism for lifting the ground state degeneracy, through the inclusion of random bonds. Similarly, spinels of the form {{\ce{AB_2X_4}}} show a number of different ordering patterns, including both an ObD and a spin liquid regime, describing diamond lattice antiferromagnets \cite{bergman2007order-by-disorder,bernier2008quantum}.

The effects of fluctuations on models and materials might be drastically altered under the combined effect of quantum and thermal fluctuations. This can happen since the two mechanisms favour magnons of different energies \cite{Henley1989,schick2020quantum}. The interplay and the distinctive effects of quantum and thermal fluctuations, although known, has been less studied in the literature, for both Ising \cite{isakov2003interplay} and Heisenberg models \cite{danu2016extended,rau2018pseudo-goldstone,khatua2023pseudo-goldstone}. In \cite{rau2018pseudo-goldstone,khatua2023pseudo-goldstone} continuous accidental degeneracies and their respective pseudo-Goldstone modes were studied, with a qualitative signature for the detection of thOBD in materials in the latter.

Lately, frustrated models have been studied in the context of Hilbert space fragmentation and quantum many body scars, as in Ref.~\cite{stephen2024ergodicity}, or shown to possess disorder-free localisation \cite{mcclarty2020disorder-free}, thus providing an avenue for connecting their study to nonthermal quantum effects \cite{turner2018weak,turner2018quantum}, nonequilibrium dynamics and simulations on quantum computers \cite{bluvstein2021controlling,ebadi2021quantum,semeghini2021probing,ebadi2022quantum,bluvstein2022a-quantum}.

In this paper, we discuss spin models whose zero-temperature ground state space is obtained from a nonlinear cellular automaton (CA) constraint. More concretely, we take an elementary CA whose update rule is a nonlinear function of the values of the cells in the previous timestep and treat trajectories of this deterministic dynamical process in 1$+$1 dimensions as defining the ground states of a spin model at zero temperature in two space dimensions. This is the general procedure we have followed in Ref.~\cite{sfairopoulos2023cellular} to construct the full list of models from all 256 elementary CA rules; in this work, we focus on the nonlinear CA rules and their respective (classical and quantum) spin models. Our approach includes three essential steps: (i) define the CA rule, (ii) find a classical spin model whose zero-temperature ground states correspond to the evolution of the respective CA, and (iii) extend the classical spin models to the quantum realm by the addition of quantum fluctuation terms in the form of a transverse field term.

We will show how the classical configuration energy can be expressed as a sum of local terms, each containing a product of (up to four) neighboring spins. For the nonlinear rules, we argue that these interactions are inherently frustrated, due to competing couplings within each local unit. In addition, we will demonstrate that the transverse field gives rise to an interplay of quantum and classical terms, quantum phase transitions and, in some cases, a quantum ObD mechanism. 

In Sec.~\ref{intro_to_CA}, we study the periodic structure of the underlying CA in the presence of periodic boundary conditions (PBC), highlighting differences from the linear CA considered in Ref.~\cite{sfairopoulos2023cellular} and their consequences for the corresponding spin models. After introducing the classical spin models in Sec.~\ref{classical_models}, we then minimally couple them to a transverse field which gives rise to nontrivial quantum effects, and study in Sec.~\ref{QPT} their ground state quantum phase transitions. In the appendix we give more detail on the periodic structure of the studied CA and some example periodic orbits.
\section{Introduction to Cellular Automata}{\label{intro_to_CA}}

Cellular automata (CA) describe a lattice of discrete cells, each hosting a degree of freedom. In general this can take values from any finite field, but here we consider the simplest scenario where it is just a binary variable. CA describe the dynamic effect of the update of these arrays of cells with time according to a given (deterministic or probabilistic) rule \cite{wolfram1983statistical,martin1984algebraic,2018probabilistic,neumann1963collected,wolfram2002a-new-kind}.

\begin{figure*}
    \centering
     \begin{subfigure}[b]{0.24\textwidth}
      \includegraphics[width=\textwidth, height=28mm]{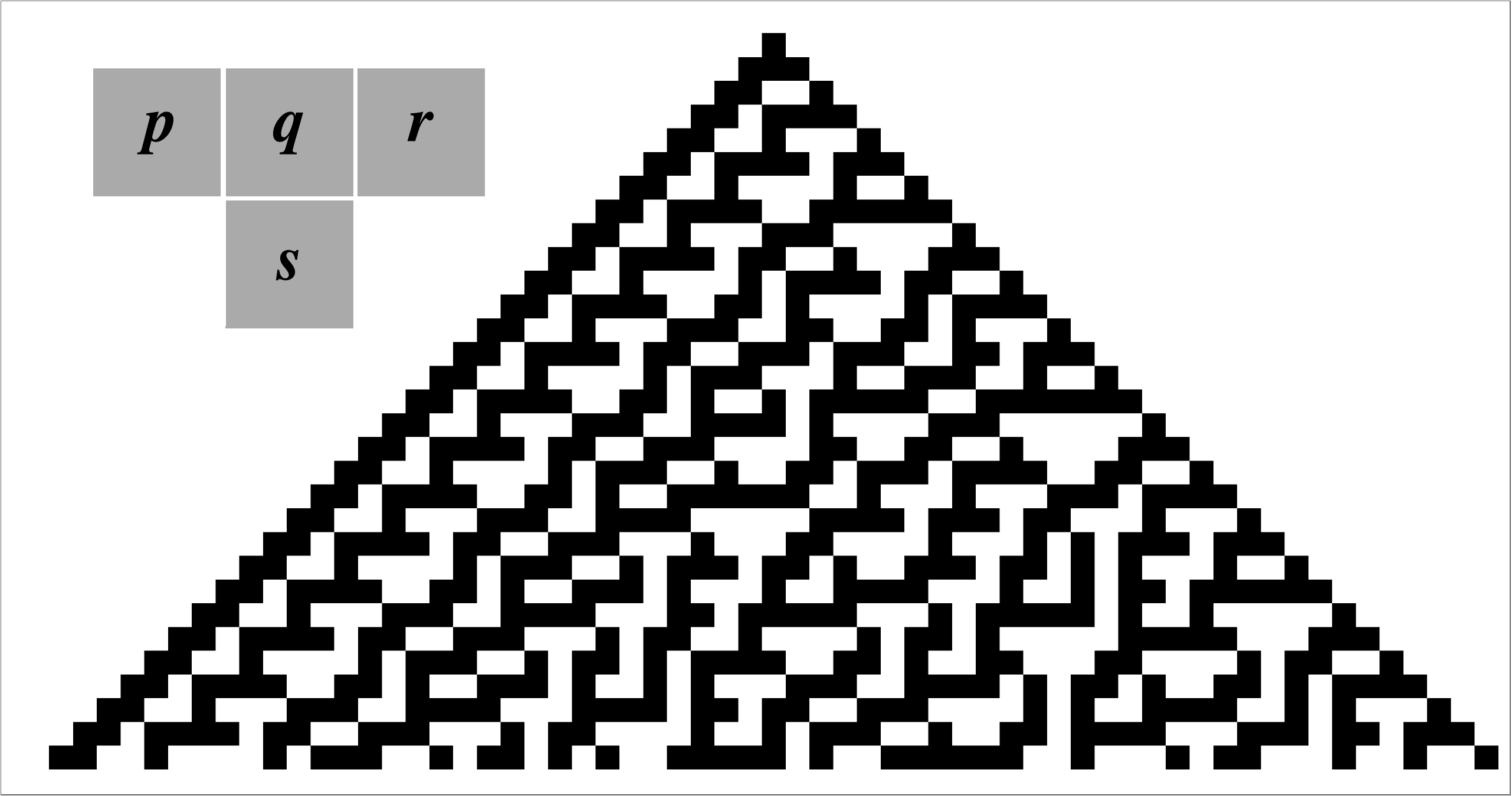}
      \caption{}
   \end{subfigure}
     \begin{subfigure}[b]{0.24\textwidth}
        \includegraphics[width=\textwidth, height=28mm]{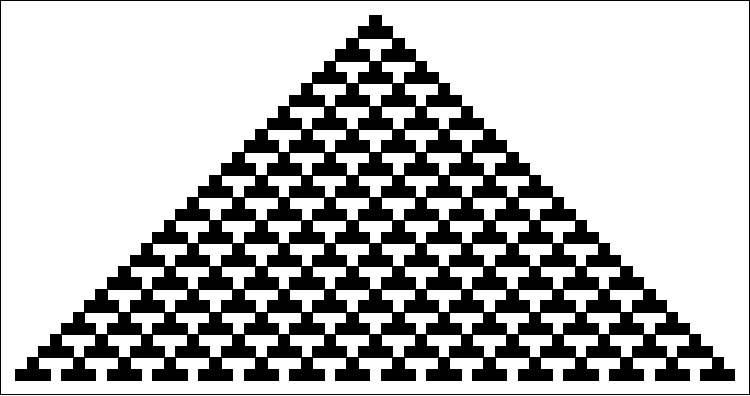}
        \caption{}
     \end{subfigure}
     \begin{subfigure}[b]{0.24\textwidth}
        \includegraphics[width=\textwidth, height=28mm]{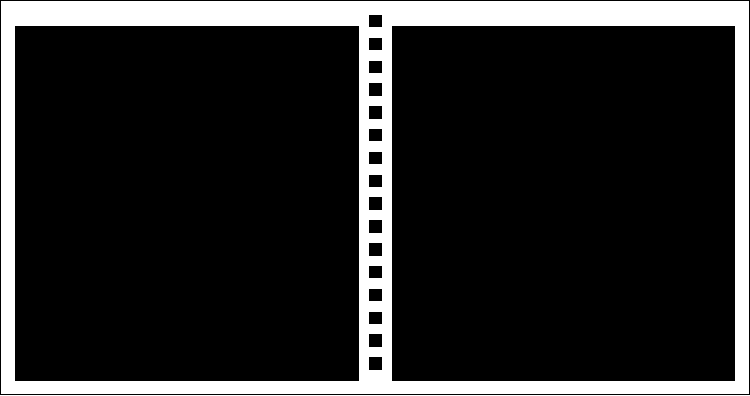}
        \caption{}
     \end{subfigure}
     \caption{
         {\bf CA with nonlinear rules.}     
         Evolution from a single occupied site for (a) Rule 30 (chaotic, non-repeating pattern), (b) Rule 54 (periodic pattern) and (c) Rule 201 (note that this starts from a single occupied site and immediately flips to a nearly all-occupied configuration). Black squares denote occupied cells of the CA ($x=1$) or spin down in the spin model ($\sigma = -1$), while white squares are unoccupied ($x = 0$) or spin up ($\sigma = 1$).
      }
     \label{fig:Rule_cycles_nonlinear_rules}
 \end{figure*}

In this work, we will consider deterministic 1D ``elementary'' CA, meaning that their update rule is local. At each timestep \(t\), the value \(s\) of a given cell is determined by a function of the value in the previous timestep of the given cell, \(q\), and of its nearest neighbors \(p\) and \(r\) [see inset of Fig.~\ref{fig:Rule_cycles_nonlinear_rules}(a)],
\begin{equation}{\label{eq:CArulepqrs}}
   s = f_n(p, q, r) \;\; \pmod 2,
\end{equation} 
where \(n\) is the rule number, or, more explicitly,
\begin{equation}
   x_{i,t+1} = f_n(x_{i-1,t}, x_{i,t}, x_{i+1,t}) \;\; \pmod 2, 
\end{equation}
where $x_{i,t} \in \{0,1\}$ is the value of site $i$ at timestep $t$.
A nonlinear CA is one where the function \(f_n\) is nonlinear in its arguments, leading to some major differences compared to the linear CA studied previously in Refs.~\cite{sfairopoulos2023boundary,sfairopoulos2023cellular}. 

Let us now describe these differences through the study of some specific examples. For this work, we will focus on three nonlinear elementary CA, Rule 30, 54 and 201. Their update rules are given by 
\begin{alignat}{2}
    &s = f_{30}(p, q, r) &&= p + q + r + qr \nonumber \\
    &s = f_{54}(p, q, r) &&= p + q + r + pr {\label{CA rules}} \\
    &s = f_{201}(p, q, r) &&= p + q + r + pr + 1 \nonumber, 
\end{alignat}
where addition is modulo 2.
Note that rules 54 and 201 are complementary \cite{wolfram2002a-new-kind}: \(f_{201}(p,q,r) = 1 - f_{54}(p,q,r) \pmod 2\).
Example trajectories with a single nonzero initial seed for the above rules are shown in Fig.~\ref{fig:Rule_cycles_nonlinear_rules}.

In this work we will focus on the study of spin models for the case of fully periodic boundary conditions (PBC). This results from the intuition gained from Ref.~\cite{sfairopoulos2023boundary}. There the triangular plaquette model was studied for the cases of open boundary conditions (OBC), but also PBC and periodic boundaries in only one dimension (PBCx). The subextensive number of the ground states in these cases obscured the numerics obtained for the identification of the quantum phase transition of the model. Similarly here, anticipating an obstruction of the same kind, we restrict to studying the given spin models only for PBC. As a result, the respective CA would need to be studied for periodic boundaries in both their space and time dimensions, thus obtaining the periodic structure of the aforementioned rules following Ref.~\cite{sfairopoulos2023cellular}.

An important difference of the nonlinear CA rules studied here, compared to the linear CA studied in Ref.~\cite{sfairopoulos2023cellular}, is the lack of either a number theoretic description for obtaining their periodic structure \cite{wolfram1983statistical} or the ability to use gaussian elimination \cite{mezard2009information}. This is due to the absence of a matrix description of the update rule \cite{wolfram1983statistical,martin1984algebraic}. In other words, the models (and the respective CA) studied in Refs.~\cite{sfairopoulos2023boundary,sfairopoulos2023cellular} consist of XORSAT instances of constrained satisfaction problems, while the nonlinear rules here of general SAT instances \cite{montanari_book}. 

The periods of rules 30, 54, and 201 for various system sizes are discussed in App.~\ref{AppendixA}, where also some small nontrivial cycles are depicted. Periodic trajectories for the specific system sizes we study below are illustrated in Figs.~\ref{fig:rule201_tiles}--\ref{fig:rule30_tiles}.
\section{Classical Spin Models from nonlinear CA}{\label{classical_models}}

In this section we consider the dynamical trajectories of the nonlinear elementary CA as ground states of two-dimensional classical spin models. Borrowing language from quantum many-body, this correspondence allows us to define (classical) parent Hamiltonians \cite{auerbach_interacting_1994} that have the chosen configurations as their ground states.

To define the spin models, we map from binary variables $x_{i,t} \in \{0,1\}$ at position \(i\) and timestep \(t\) of the CA to Ising spin variables $\sigma_{i,t} = 1 - 2x_{i,t} = \pm 1$ at position $(i,t)$ in 2D space. For a general rule, we define a classical energy function whose ground states are given by CA rule \(n\), taking the form 
\begin{align}
    \begin{split}
        E_n &= - \sum_{\{p, q, r, s\}}^N \left( 2 \delta_{s, f_n(p, q, r)} - 1  \right) \\
            &= - \sum_{i,j}^{L,M} \left( 2 \delta_{x_{i,j+1}, f_n(x_{i-1,j}, x_{i,j}, x_{i+1,j})} - 1 \right) \\
            &= - \sum_{\{p, q, r, s\}}^N d_{n}(p, q, r, s).
    \end{split}{\label{definition of energy functions}}
\end{align}
In the first line, we have used notation coming from CA, with the ``plaquette'' (i.e., set of sites subject to the constraint) \(p\), \(q\), \(r\) and \(s\) arranged as in Eq.~\ref{eq:CArulepqrs}. In the second line we have used $\{i,j\}$ to denote the summation over the two space dimensions, of size $L$ and $M$, respectively, of the spin model. The Kronecker delta in both cases enforces the given rule.

In the third line of Eq.~\ref{definition of energy functions} we have reexpressed the constraint into the form of a local defect variable $d_{n}(p, q, r, s)$ \cite{garrahan2000glassiness,garrahan2002glassiness} taking the values $\pm 1$, with $-1$ indicating the presence of a defect. This locally constrains each of the $N = L \times M$ plaquettes according to the corresponding rule. In this description, defects of a given rule form fractal patterns for infinite systems, similarly to the Sierpinski triangles for the triangular plaquette model \cite{garrahan2000glassiness}. 

We now rewrite Eq.~\ref{definition of energy functions} in a more conventional way for a spin model, by mapping each term into an explicit interaction between the spins. For a linear CA, this gives a product of spins on the corresponding sites; e.g., for Rule 60, $s = p + q \pmod 2$, the spin interaction term is $-\sigma_p \sigma_q \sigma_s$ \cite{newman1999glassy,sfairopoulos2023boundary}. For a nonlinear CA, it gives rise to controlled $Z$ gates: a product of two binary terms in \(f_n\) maps to a CZ gate, a three-site product to a CCZ gate, and so on. For example, given a CA that has a term $pq$, this would correspond to $pq \to \text{CZ}_{pq} = (1 + \sigma_p + \sigma_q - \sigma_p \sigma_q)/2$.

Applying this procedure to Rules 30, 54 and 201, we obtain
\begin{align}{\label{non-onsite_rules_equations1}}
    E_{30} &=  -\sum_{\{p, q, r, s\}}^N d_{30}(p, q, r, s) =  -\sum_{\{p, q, r, s\}}^N \sigma_p \sigma_q \sigma_r CZ_{qr} \sigma_s \\
    E_{54} &=  -\sum_{\{p, q, r, s\}}^N d_{54}(p, q, r, s) =  -\sum_{\{p, q, r, s\}}^N \sigma_p \sigma_q \sigma_r CZ_{pr} \sigma_s, {\label{non-onsite_rules_equations2}}
\end{align}
and \(E_{201}= - E_{54}\).
By decomposing the CZ gate, Eq.~\ref{non-onsite_rules_equations1} and Eq.~\ref{non-onsite_rules_equations2} can be expressed as a sum of interactions corresponding to linear CA rules,
\begin{align}{\label{rules_equations1}}
    &E_{30} = \frac{1}{2} \left( - E_{240} + E_{60} + E_{90} + E_{150}  \right) \\
    &E_{54} = \frac{1}{2} \left( - E_{204} + E_{60} + E_{102} + E_{150}  \right) {\label{rules_equations3}},
\end{align}
where
\begin{align}
    &E_{240}\;   = - \sum_{\{p, s\} \in \Isingdiagonal}^N       \sigma_p \sigma_s     \label{m240}    
    \\
    &E_{204} \;  = - \sum_{\{q, s\} \in \Isingvertical}^N       \sigma_q \sigma_s     \label{m204}  
    \\
    &E_{60} \, \, \;= - \sum_{\{p, q, s\} \in \trigonright}^N    \sigma_p \sigma_q \sigma_s    \label{m60}    
    \\
    &E_{90} \; \, \,= - \sum_{\{p, r, s\} \in \trigonmiddle}^N    \sigma_p \sigma_r \sigma_s     \label{m90}    
    \\
    &E_{102} \;     = - \sum_{\{q, r, s\} \in \trigonleft}^N    \sigma_q \sigma_r \sigma_s     \label{m102}    
    \\
    &E_{150} \;     = - \sum_{\{p, q, r, s\} \in \tetrapleuro}^N \sigma_p \sigma_q \sigma_r \sigma_s
    \label{m150}    
\end{align}
are classical energy functions defined in the same way \cite{newman1999glassy,sfairopoulos2023boundary}. The above classical energy terms correspond to all the allowed plaquette terms which come from linear CA constraints from the elementary CA.

The two complementary expressions for \(E_n\), in terms of defect variables \(d_n\) and in terms of products of spins, elucidate the frustration in the spin models. From the defect description of the parent Hamiltonians (third line of Eq.~\ref{definition of energy functions}) we infer that each elementary plaquette takes values $\pm 1$, while, at the same time, these terms are equal to the sum of four local plaquette terms (Eqs.~\ref{rules_equations1} and \ref{rules_equations3}). One of the four terms must therefore take the value \(+1\) in each ground state, indicating the presence of frustrated interactions.
\section{Quantum models and Quantum Phase transitions}{\label{QPT}}

In this section, we study the quantum version of these models by minimally coupling them to a transverse field, 
\begin{equation}{\label{quantum models}}
    H_{30/54/201} = J E_{30/54/201} - h \sum_{i,j}^{L,M} X_{i,j},
\end{equation}
where in the classical energy terms we use the basis of the Pauli $Z$-operator for the spins, $\sigma_{i,j} \rightarrow Z_{i,j}$, where $X_{i,j}$ and $Z_{i,j}$ are the Pauli matrices acting on site ${i,j}$.

The purpose of this section is to examine the interplay of quantum fluctuations, as introduced by the transverse field term in Eq.~\ref{quantum models}, and the classical frustrated interactions. For this reason, in Sec.~\ref{qObD}, we apply degenerate perturbation theory to study the models for small system sizes in the limit of small \(h/J\), while also inferring their behavior in the thermodynamic limit where possible. In Sec.~\ref{QPTs} we verify these claims and extend our results to larger \(h/J\) using numerical simulations. We use a combination of exact diagonalization, matrix product states and quantum Monte Carlo techniques which also allow us to identify indications of a quantum phase transition for all the models studied.

\subsection{Quantum Order-by-Disorder}{\label{qObD}}

We first utilize degenerate perturbation theory \cite{schrieffer1966relation,bravyi2011schriefferwolff,slagle2017fracton,sakurai2020modern} to treat the nonlinear models on a case-by-case basis and for specific small system sizes. We then try to generalize to the thermodynamic limit (if any) and to other rules.

\subsubsection{Degenerate perturbation theory}
\label{DegeneratePerturbationTheory}

Consider a Hamiltonian, $H = H^0 + h H^1$, where \(H^0\) has degenerate ground states labeled by ${\mu}$, i.e., $H^0 \ket{g_{\mu}} = E^0 \ket{g_{\mu}}$, and $H^1$ is the perturbing potential with $h$ much smaller than the energy gap of the system. The index $\mu$ refers to the ground state degeneracy. The projector $P$ projects onto the ground state manifold, so that 
\begin{equation}
   H^0 P = P H^0 = E^0 P.
\end{equation}
We define an effective Hamiltonian \(H_{\text{eff}}\) through a canonical transformation given by the antihermitian operator $S$. Collecting terms order by order, 
\begin{align}
   H_{\text{eff}} &= e^{S} H e^{-S} = H + [S, H] + \frac{1}{2!} \bigl[S, [S, H]\bigr] + \cdots \nonumber \\
   &= H^0 + h \left( \bigl[S^1, H^0\bigr] + H^1 \right) + O\left(h^2 \right),
\end{align}
where $S = \sum_{n=1}^{\infty} h^n S_n$. We require that 
$\bigl[P, H_{\text{eff}}\bigr] = 0$, which ensures the decoupling of the ground and excited state manifolds; see \cite{schrieffer1966relation,bravyi2011schriefferwolff,slagle2017fracton,sakurai2020modern}. 

The first and second order terms in this expansion are 
\begin{align}
   H_{\text{eff}}^{(1)} &= PH^1P \\
   H_{\text{eff}}^{(2)} &= P H^1 \frac{1-P}{E^0 - H^0} H^1 P .
   \label{EqHeff2}
\end{align}
In our case, we use $H^0 = J E_{30/54/201}$ and $H^1 = - \sum_{i,j} X_{i,j}$.

The first-order correction always vanishes for any of the models studied (and in general of this CA class), because the rules are deterministic. The single spin flip caused by \(H^1\) always takes the system out of the set of CA trajectories (assuming PBC are imposed \footnote{Note that for OBC the situation is drastically different and all models will acquire corrections to first order due to the freedom of selecting a number of spins of the last row, while also due to the update rule of the given CA itself. More precisely, a classical ground state spin term on the east boundary will get updated according to the given CA rule, say Rule 54. For, example, for the form $\downarrow \uparrow \uparrow \; \; \rightarrow \; \; \downarrow$ while at the same time $\downarrow \uparrow \downarrow \;\; \rightarrow \;\; \downarrow$. Similarly, the south-east boundary will possess one spin which will not take part in any interaction term. This shows that locally at the boundary first order degenerate perturbation theory will couple the classical zero temperature ground states, thus giving a nonzero contribution.}) and hence out of the ground state manifold.

The second order perturbation therefore gives the first nontrivial correction to the energy levels. Using Eq.~\ref{EqHeff2}, it involves the matrix elements
\begin{equation}{\label{second order PT}}
\braket{g|H_{\text{eff}}^{(2)}|g'} = - \sum_{e} \frac{\braket{g | H^1 | e} \braket{e | H^1 | g'}}{\Delta E}
\end{equation}
where \(\ket{g}\) and \(\ket{g'}\) are ground states, $\ket{e}$ is an excited state, and $\Delta E = E_{e} - E^0 > 0$ is the energy difference between the two. Each term in the sum is only nonzero if \(\ket{e}\) differs by a single spin flip from both \(\ket{g}\) and \(\ket{g'}\).

\begin{figure}
    \centering
    \begin{subfigure}[b]{0.07\textwidth}
        \includegraphics[width=10mm, height=10mm]{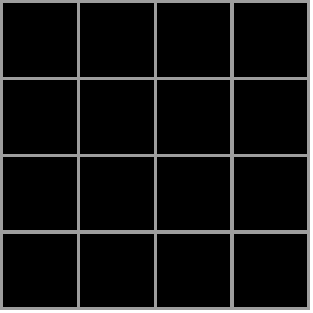}
        \caption{}
     \end{subfigure}
     \begin{subfigure}[b]{0.07\textwidth}
        \includegraphics[width=10mm, height=10mm]{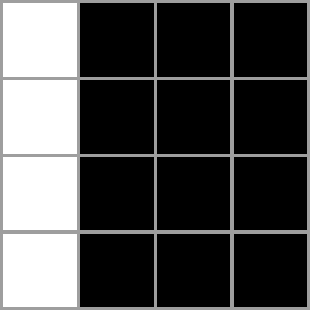}
        \caption{}
     \end{subfigure}
     \begin{subfigure}[b]{0.07\textwidth}
        \includegraphics[width=10mm, height=10mm]{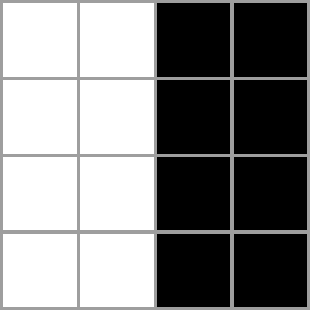}
        \caption{}
     \end{subfigure}
     \begin{subfigure}[b]{0.07\textwidth}
        \includegraphics[width=10mm, height=10mm]{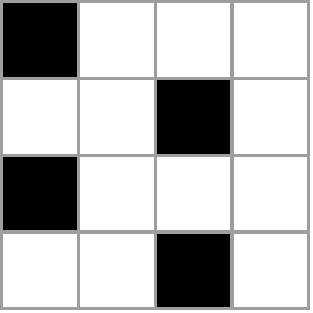}
        \caption{}
     \end{subfigure}
     \caption{The ground states (or, equivalently, periodic CA trajectories) for Rule 201 for a $4 \times 4$ system size. As in Fig.~\ref{fig:Rule_cycles_nonlinear_rules}, black and white squares are, respectively, down ($\sigma = -1$) and up ($\sigma = +1$) spins (or occupied and unoccupied cells of the CA). Ground states which are obtained by translations of these states are omitted.}
     \label{fig:rule201_tiles}
 \end{figure}

\begin{figure}
    \centering
     \begin{subfigure}[b]{0.07\textwidth}
        \includegraphics[width=10mm, height=5mm]{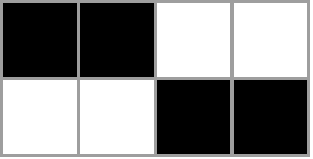}
        \caption{}
     \end{subfigure}
     \begin{subfigure}[b]{0.07\textwidth}
        \includegraphics[width=10mm, height=5mm]{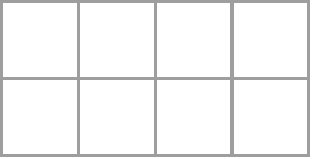}
        \caption{}
     \end{subfigure}
     \caption{Similar to Fig.~\ref{fig:rule201_tiles} for the ground states of the Rule 54 for a $4 \times 2$ system size.}
     \label{fig:rule54_tiles}
 \end{figure}

\begin{figure}
   \centering
    \begin{subfigure}[b]{0.07\textwidth}
       \includegraphics[width=10mm, height=10mm]{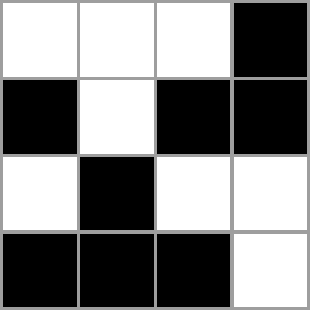}
       \caption{}
    \end{subfigure}
    \begin{subfigure}[b]{0.07\textwidth}
       \includegraphics[width=10mm, height=10mm]{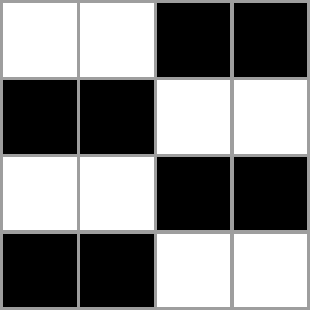}
       \caption{}
    \end{subfigure}
    \begin{subfigure}[b]{0.07\textwidth}
      \includegraphics[width=10mm, height=10mm]{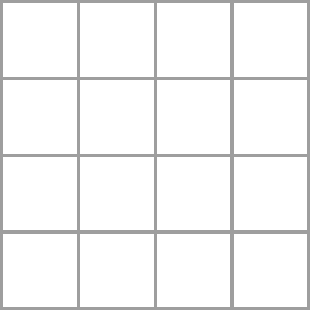}
      \caption{}
   \end{subfigure}
    \caption{Similar to Fig.~\ref{fig:rule201_tiles} for the ground states of the Rule 54 for a $4 \times 4$ system size.}
    \label{fig:rule54_tiles_4x4}
\end{figure}

\begin{figure}
    \centering
    \begin{subfigure}[b]{0.07\textwidth}
        \includegraphics[width=10mm, height=20mm]{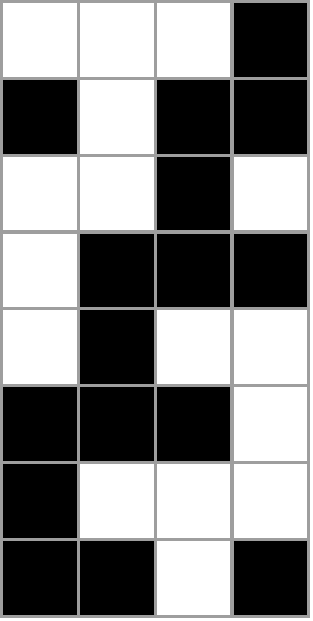}
        \caption{}
     \end{subfigure}
     \begin{subfigure}[b]{0.07\textwidth}
        \includegraphics[width=10mm, height=20mm]{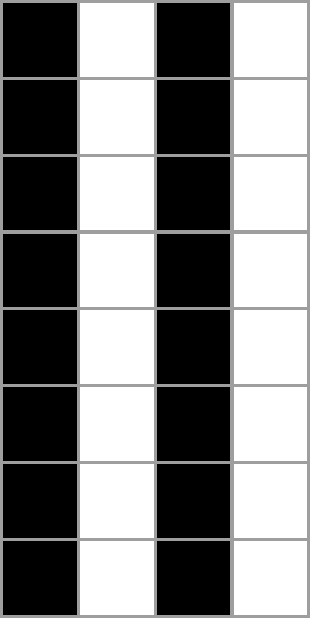}
        \caption{}
     \end{subfigure}
     \begin{subfigure}[b]{0.07\textwidth}
        \includegraphics[width=10mm, height=20mm]{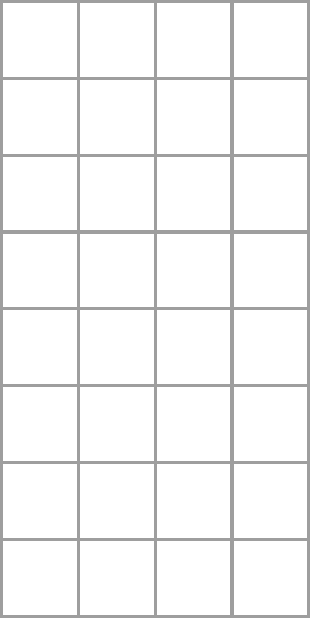}
        \caption{}
     \end{subfigure}
     \caption{Similar to Fig.~\ref{fig:rule201_tiles} for the ground states of the Rule 30 for a $4 \times 8$ system size.}
     \label{fig:rule30_tiles}
\end{figure}

\begin{figure}
   \centering
   \includegraphics[width=\linewidth]{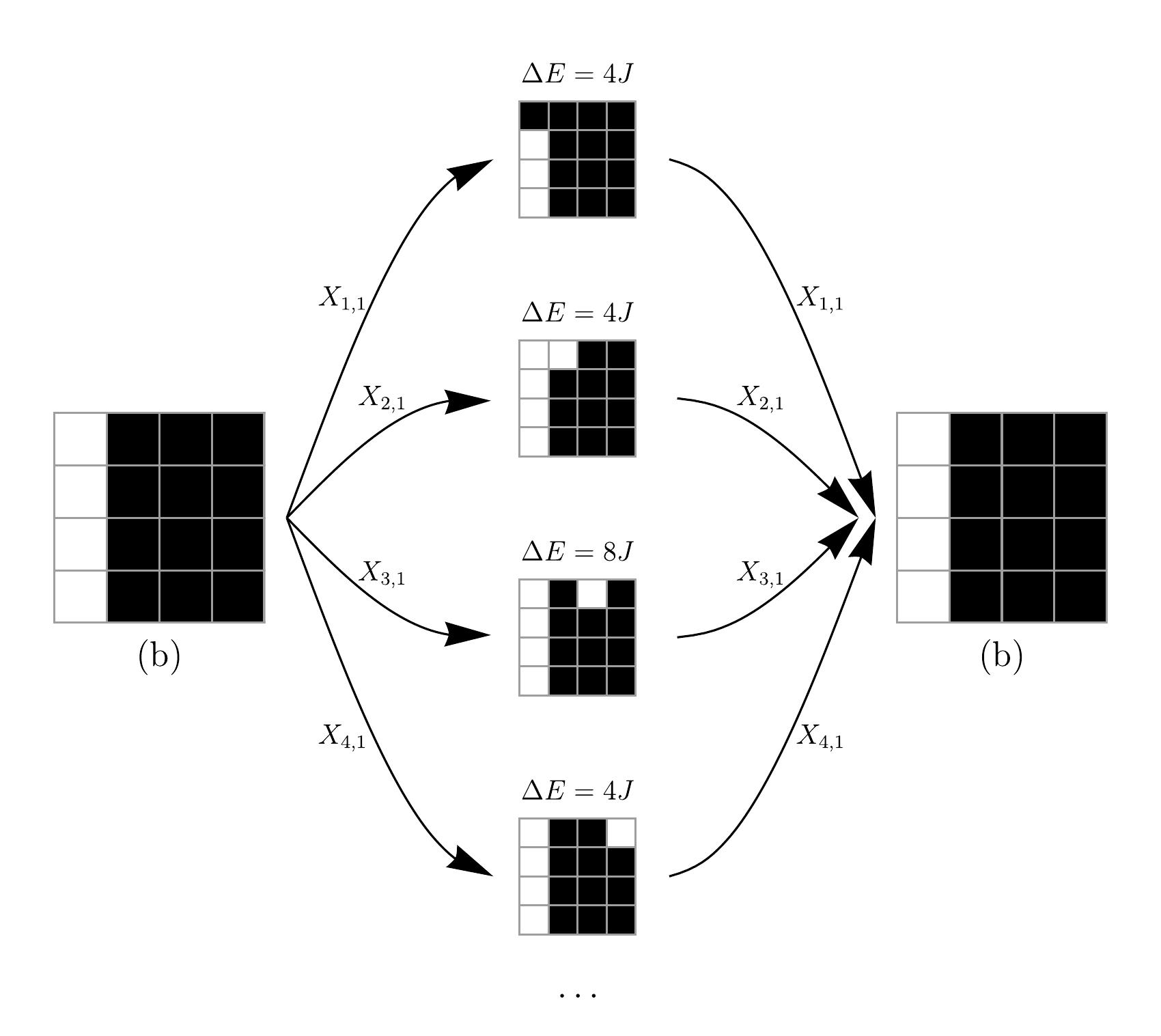}
   \caption{A sketch of the second order processes contributing to perturbation theory for the states in panel (b) of Fig.~\ref{fig:rule201_tiles}. The processes shown involve flipping the spins of the first row. Flipping the spins of the subsequent rows will lead to results analogous to the ones shown.}
    \label{fig:perturbation_theory}
\end{figure}

In general, these matrix elements might connect different ground states, but such off-diagonal terms are in fact strongly restricted at all orders in perturbation theory. Nonzero \(\braket{g|H_{\text{eff}}^{(n)}|g'}\) for \(\ket{g} \neq \ket{g'}\) occurs due to sequences of \(n\) spin flips that pass through intermediate excited states before returning to the ground-state manifold \cite{sakurai2020modern}, and hence requires two different CA trajectories to differ on at most \(n\) sites. It can be seen from Figs.~\ref{fig:rule201_tiles}--\ref{fig:perturbation_theory} that this does not occur for \(n=2\) in the small systems we consider, and so \(\braket{g|H_{\text{eff}}^{(2)}|g'} \neq 0\) only for \(\ket{g} = \ket{g'}\). (There are exceptions for even smaller sizes, but they do not generalize to larger lattices and so we exclude them from our analysis.)

Furthermore, it is a general property of spin models (with PBC) derived from deterministic CA that any two different ground states differ by a number of spin flips that scales at least with the linear system size. To see this, consider flipping a single spin and then looking at the row above. Because the CA is deterministic, at least one of the three relevant spins in this row (corresponding to the previous timestep) must also be flipped to give a valid trajectory. Repeating this argument and using the periodic boundary conditions, there must be at least one flipped spin in each row. Off-diagonal terms in \(H_{\text{eff}}\) therefore appear only at an order \(n\) that grows at least subextensively with system size.



\subsubsection{Rule 201}
\label{qObD201}

Let us now study the specific cases of Rules 30, 54 and 201 in turn. We start with Rule 201 and the system size $L\times M = 4\times 4$. There are 13 classical ground states of the model for this system size, of which the 4 representatives (up to translations) are shown in Fig.~\ref{fig:rule201_tiles}. As noted above, an infinitesimal transverse field splits the classical ground states to second order in perturbation theory, and each nonzero term in the sum in Eq.~\ref{second order PT} can be interpreted as a process involving a sequence of spin flips.

The diagonal matrix elements for the states of panel (b) of Fig.~\ref{fig:rule201_tiles} correspond to the processes sketched in Fig.~\ref{fig:perturbation_theory}. There are 4 such ground states, related by translation symmetry, and each receives a nonzero contribution from every excited state \(\ket{e}\) that is related to it by a single spin flip. Of these \(L\times M=16\) excited states, 12 have two excited plaquettes and so \(\Delta E = +4J\), while the remaining 4 have four excited plaquettes and so \(\Delta E = +8J\). The diagonal matrix element is therefore
\begin{equation}
\Braket{ g^{(b)} | H_{\text{eff},201}^{(2)} | g^{(b)} } =-N \left(\frac{3}{4} \frac{1}{4J} + \frac{1}{4} \frac{1}{8J}
\right).
\end{equation}
where \(N = LM = 16\). Similar calculations for the other panels in Fig.~\ref{fig:rule201_tiles} give
\begin{align}
\label{EqE2_201}
   \begin{split}
   E^{(2)}_{201, (a)} = \Braket{ g^{(a)} | H_{\text{eff},201}^{(2)} | g^{(a)} } &=-\frac{N}{4J} \\
   E^{(2)}_{201, (b)} = \Braket{ g^{(b)} | H_{\text{eff},201}^{(2)} | g^{(b)} } &=
   -\frac{7N}{32J} \\
   E^{(2)}_{201, (c)} = \Braket{ g^{(c)} | H_{\text{eff},201}^{(2)} | g^{(c)} } &= -\frac{3N}{16J} \\
   E^{(2)}_{201, (d)} = \Braket{ g^{(d)} | H_{\text{eff},201}^{(2)} | g^{(d)} } &= -\frac{5N}{32J}, 
   \end{split}
\end{align}
which (because all off-diagonal elements vanish to this order) give the second-order corrections to the energy eigenvalues. Note that the matrix is 13-dimensional in this case, but the energy eigenvalues $E^{(2)}_{201, (b)}$ to $E^{(2)}_{201, (d)}$ remain fourfold degenerate due to translational invariance. Because
\begin{equation}
   E^{(2)}_{201, (a)} < E^{(2)}_{201, (b)} <E^{(2)}_{201, (c)} < E^{(2)}_{201, (d)}.
\end{equation}
the quantum effects select the (single) trivial ground state of panel (a) in Fig.~\ref{fig:rule201_tiles}.

Now, we want to address how this perturbation theory would apply to this model for larger systems and in the thermodynamic limit. As explained in Sec.~\ref{DegeneratePerturbationTheory}, different classical ground states differ from each other by a number of spin flips that is at least linear in the linear system size, and hence the perturbation theory would lead to nonzero off-diagonal corrections in an order which scales at least with the linear system size. 

At the same time, we can infer for Rule 201 which classical ground states receive the largest (most negative) energy corrections based on the above calculations. For a configuration with all spins down, such as in Fig.~\ref{fig:rule201_tiles}(a), every excited state \(\ket{e}\) has a single up spin, giving two excited plaquettes and hence \(\Delta = +4J\). The second-order energy shift is therefore \(E_{201}^{(2)} = -\frac{N}{4J}\), as in Eq.~\ref{EqE2_201} for the \(4\times 4\) case. All other classical ground states contain up spins, which increase \(\Delta E\) and hence give a smaller correction.

We thus conclude that, for all system sizes, quantum fluctuations select the state with all spins down, at least for sufficiently small transverse fields. (The true quantum ground state is not equal to this classical configuration, because the unitary transformation \(e^{S}\) gives a superposition including other states.) This is an example of diagonal ObD, where diagonal elements of \(H_{\text{eff}}\), induced by quantum fluctuations, select a particular configuration. In this case, the ordering mechanism does not break of any symmetry of the model, since the selected configuration is spatially symmetric and the classical energy function does not have spin-inversion symmetry.

\subsubsection{Rule 54}
\label{qObD54}

We now consider $H_{54}$, starting with a $4 \times 2$ system size \footnote{Note that we start our study of Rule 54 from the system with size $4 \times 2$, while for Rule 201 we started from the $4 \times 4$. This is because the system size $4 \times 2$ for Rule 201 leads to off-diagonal terms in the second order term of the degenerate perturbation theory. These cross-terms between the different ground states are observed only for small system sizes. In the thermodynamic limit, we expect no off-diagonal terms to second order.}. A small transverse field again gives nontrivial corrections to the classical ground states to second order in perturbation theory. There are 5 ground states in this case, whose 2 representatives are shown in Fig.~\ref{fig:rule54_tiles}. Their energy corrections are
\begin{align}
   \begin{split}
   E^{(2)}_{54, (a)} = \Braket{ g^{(a)} | H_{\text{eff},54}^{(2)} | g^{(a)} } & = -\frac{3N}{16J} \\
   E^{(2)}_{54, (b)} = \Braket{ g^{(b)} | H_{\text{eff},54}^{(2)} | g^{(b)} } & = -\frac{N}{8J},
   \end{split}
\end{align}
respectively. (All off-diagonal matrix elements are again zero to second order.) Because \(E^{(2)}_{54, (a)} < E^{(2)}_{54, (b)}\), the four states represented by panel (a) of Fig.~\ref{fig:rule54_tiles} and related to each other by translations acquire the largest energy correction. Quantum fluctuations therefore select states that break the translation symmetry, in contrast to Rule 201.

We show the $4 \times 4$ system size in Fig.~\ref{fig:rule54_tiles_4x4}. There are two classes of states, (b) and (c), that are tilings of the states in Fig.~\ref{fig:rule54_tiles}. Their degeneracy is lifted at fourth order in perturbation theory.

For larger system sizes, we expect many more sets of states which are related under translations to contribute to the classical ground state manifold. These all receive energy corrections that are larger than those for the all-up states as in panel (b). This follows from the same argument as for Rule 201, but reversed; here instead of counting $\downarrow$ spins, we count $\uparrow$ spins. This is to be expected, since rules 54 and 201 are the same up to a sign change of the classical energy term. As a result, from the second order perturbation theory we expect the selection of a set of configurations that come from translationally asymmetric ground states, and which are not split at any order in perturbation theory due to their equivalence under translations. In the thermodynamic limit, we therefore expect that the ground state selection would involve the spontaneous breaking of the translation symmetries (TSSB).

Note that although this argument implies that the lowest-energy state is not the symmetric all-up configuration, and hence that there is TSSB, it does not determine which states are selected. It therefore does not allow us to predict the pattern of spatial order, or how this depends on system size.

\subsubsection{Rule 30}
\label{qObD30}

Finally, we discuss $H_{30}$. For this CA rule, we conjecture (but do not have a proof \footnote{For extensive analyses of Rule 30, see \cite{wolfram2002a-new-kind}.}) that there is a single configuration compatible with PBC for any odd \(L\) and three for any even \(L\). For the $4 \times 8$ system size shown in Fig.~\ref{fig:rule30_tiles}, we find
\begin{align}
   \begin{split}
   E^{(2)}_{30, (a)} = \Braket{ g^{(a)} | H_{\text{eff},30}^{(2)} | g^{(a)} } & = -\frac{17N}{96J} \\
   E^{(2)}_{30, (b)} = \Braket{ g^{(b)} | H_{\text{eff},30}^{(2)} | g^{(b)} } & = -\frac{3N}{16J} \\
   E^{(2)}_{30, (c)} = \Braket{ g^{(c)} | H_{\text{eff},30}^{(2)} | g^{(c)} } & = -\frac{N}{8J},
   \end{split}
\end{align}
respectively, with 
\begin{equation}
   E^{(2)}_{30, (b)} < E^{(2)}_{30, (a)} < E^{(2)}_{30, (c)}.
\end{equation}
We observe a similar pattern of diagonal ObD with additional TSSB, as for Rule 54. However, in this case we have no constructive arguments for the extrapolation to larger system sizes.

\subsubsection{Comparison of nonlinear rules}

Finally, we summarize the common features of perturbation theory for these three models and the differences. All models show a splitting of their energy eigenvalues to second order in degenerate perturbation theory. All the off-diagonal entries of this matrix are zero to this order. The classical ground states are not connected to each other in any finite order in perturbation theory in the thermodynamic limit. For both Rule 201 and Rule 54 we can argue in favour of the presence of a diagonal ObD. For the former, a single state acquires the largest energy correction, thus favouring a symmetric ground state. For the latter, a set of states get the same energy corrections and thus a ground state degeneracy is expected. However, the specific features of these states in the thermodynamic limit cannot be determined by these methods.

For Rule 30, a splitting of the energy eigenvalues to second order in degenerate perturbation theory is also encountered. However, larger system sizes do not preclude the splitting of the energy levels to a higher order in perturbation theory. At the same time, no constructive arguments exist to argue for which matrix elements will give rise to the lowest energy corrections. This aligns well with the chaotic nature of the CA Rule 30, as explained in detail in Ref.~\cite{wolfram2002a-new-kind}.

\subsection{Quantum Phase Transitions}{\label{QPTs}}

In this section we study the models of Sec.~\ref{qObD} through numerical simulations. Our goal is to verify and go beyond perturbation theory, based on calculations on larger system sizes. For this reason, we show results from numerical simulations based on exact diagonalization (ED) for limited system sizes \cite{sandvik2010computational}, and for larger system sizes from matrix product state (MPS) methods \cite{schollwock2011the-density-matrix,stoudenmire2012studying,fishman2022the-itensor,fishman2022the-itensor2} (with bond dimensions up to 1000), and continuous-time quantum Monte Carlo (ctQMC) simulations \cite{beard1996simulations,krzakala2008path-integral,mora2012transition,causer2024rejection-free} (with inverse temperature at least $\beta=100$).

\begin{figure*}[t]
    \begin{subfigure}[b]{0.24\textwidth}
       \includegraphics[width=\textwidth]{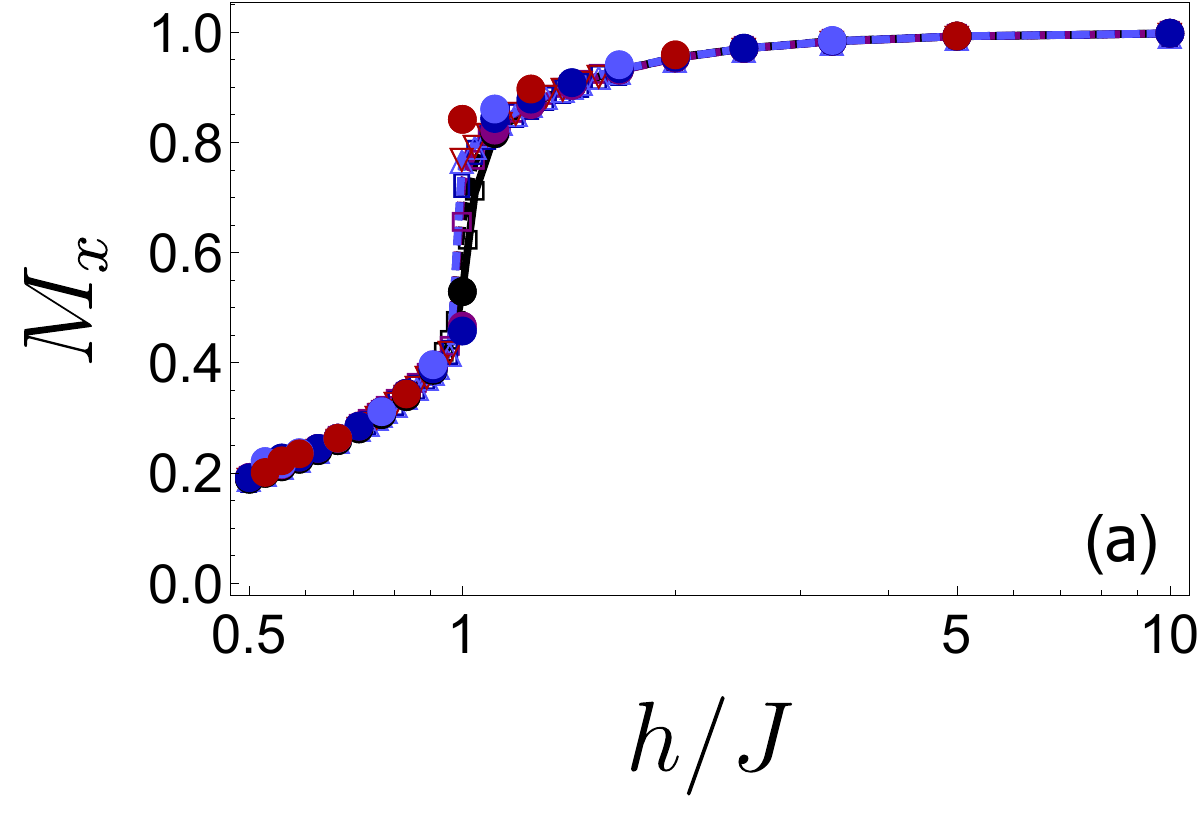}
    \end{subfigure}
    \begin{subfigure}[b]{0.24\textwidth}
       \includegraphics[width=\textwidth]{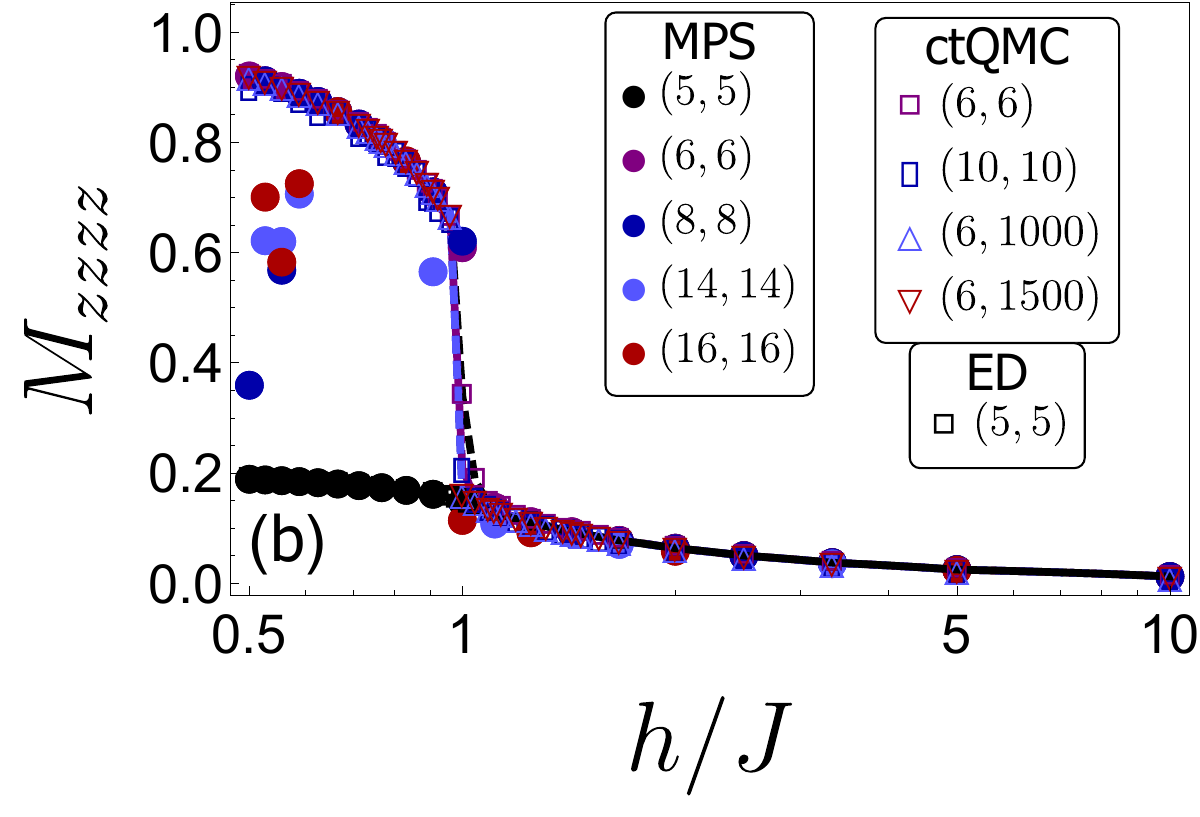}
    \end{subfigure}
    \begin{subfigure}[b]{0.24\textwidth}   
       \includegraphics[width=\textwidth]{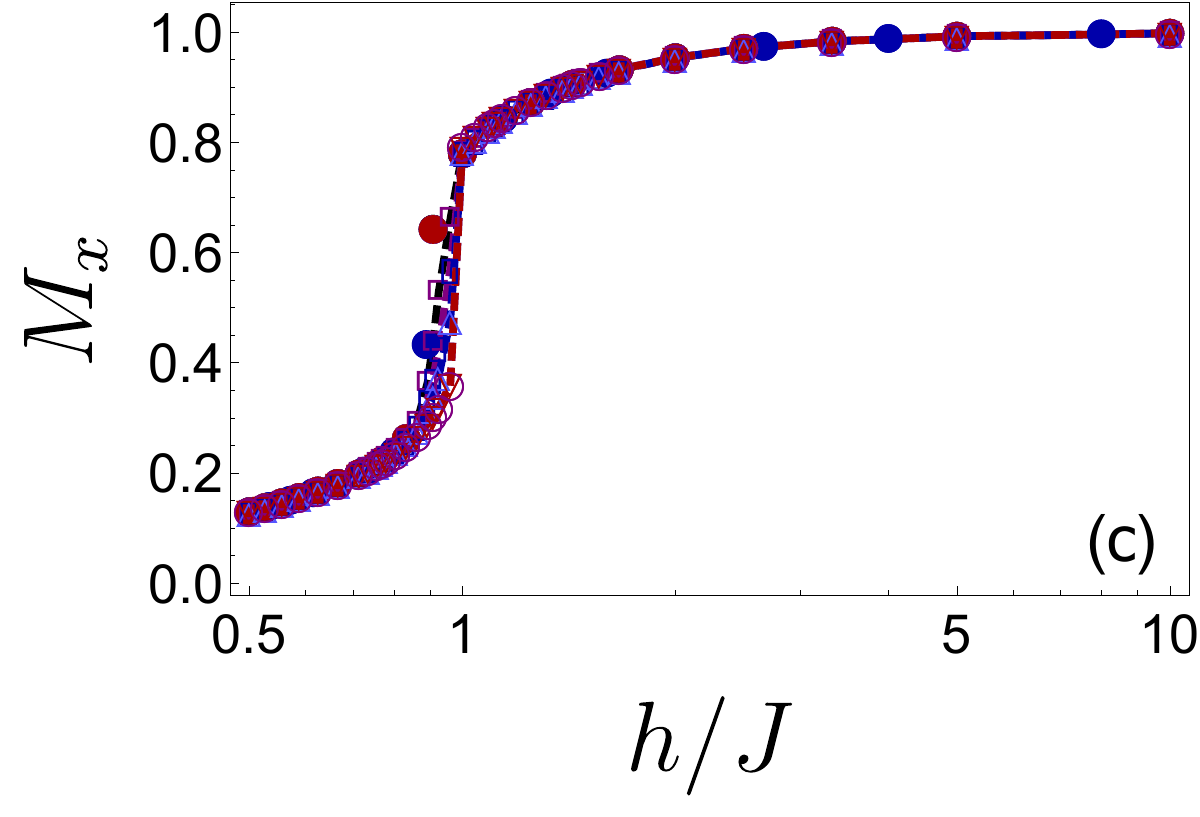}
    \end{subfigure}
    \begin{subfigure}[b]{0.24\textwidth}   
       \includegraphics[width=\textwidth]{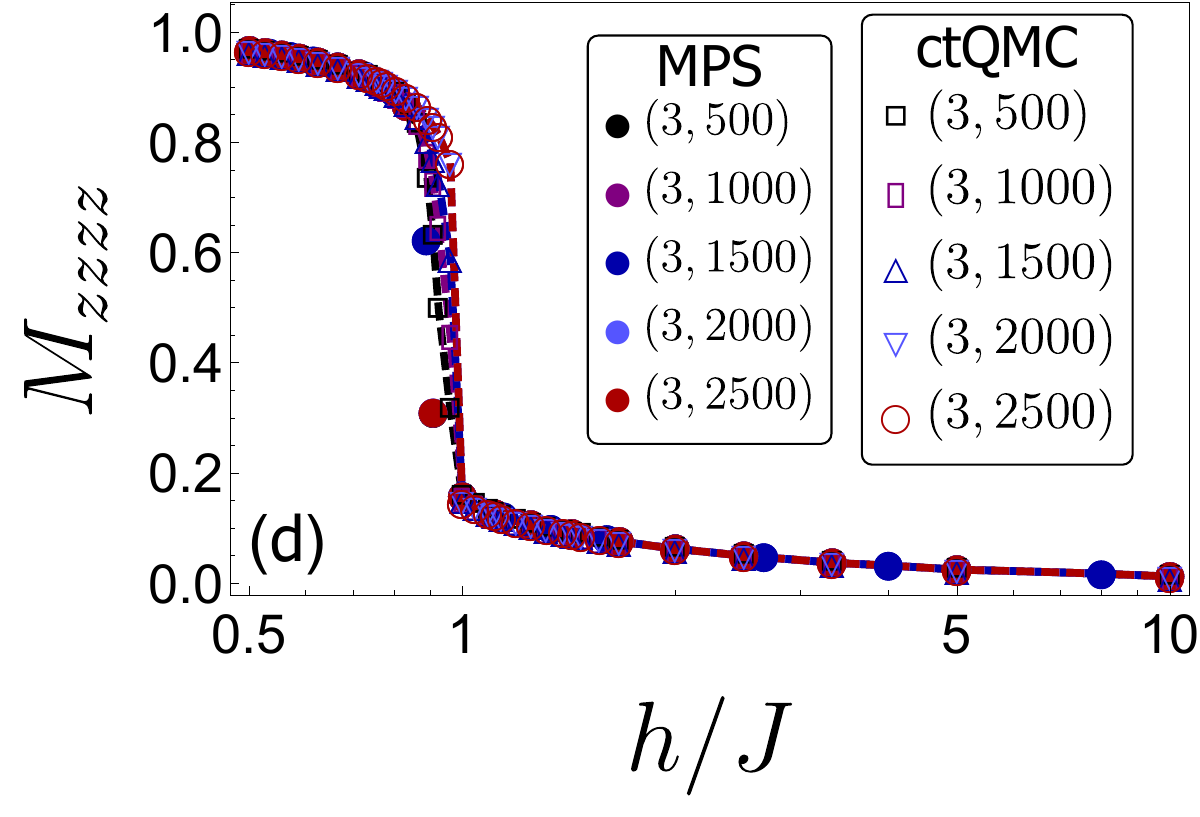}
    \end{subfigure}
    \caption{
       {\bf Quantum phase transition of $H_{30}$ for PBC.}
       (a) Transverse magnetisation $M_x$ as a function of $J$, for square system sizes $L \times L$ from numerical MPS (filled symbols) and square and rectangular sizes from ctQMC simulations (empty symbols). Exact diagonalization data shown as empty symbols and a continuous interpolating function (black).
       (b) Four-spin interaction magnetisation similar to (a).
       (c,d) Same with (a, b) for thin strip system sizes.
       }
     \label{fig:quantum rule 30 PBC}
 \end{figure*}
 
  \begin{figure*}[t]
    \begin{subfigure}[b]{0.24\textwidth}
       \includegraphics[width=\textwidth]{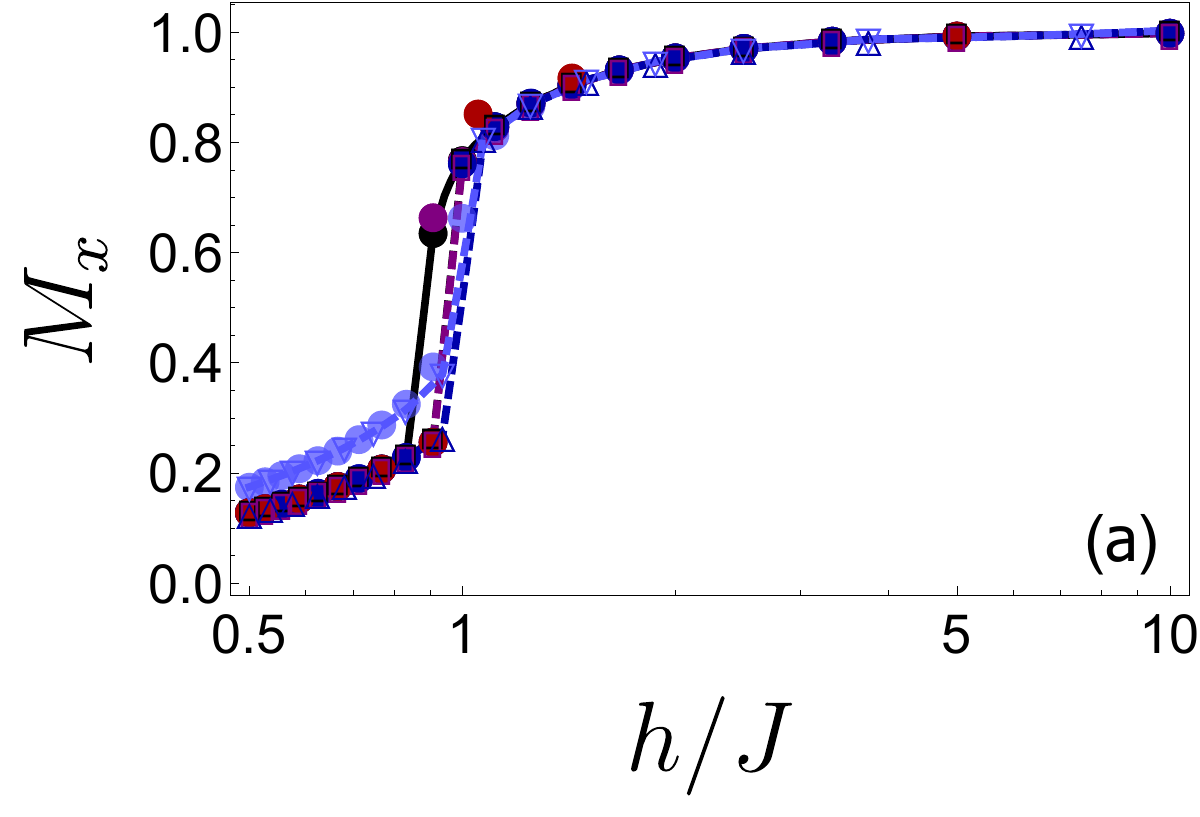}
    \end{subfigure}
    \begin{subfigure}[b]{0.24\textwidth}
       \includegraphics[width=\textwidth]{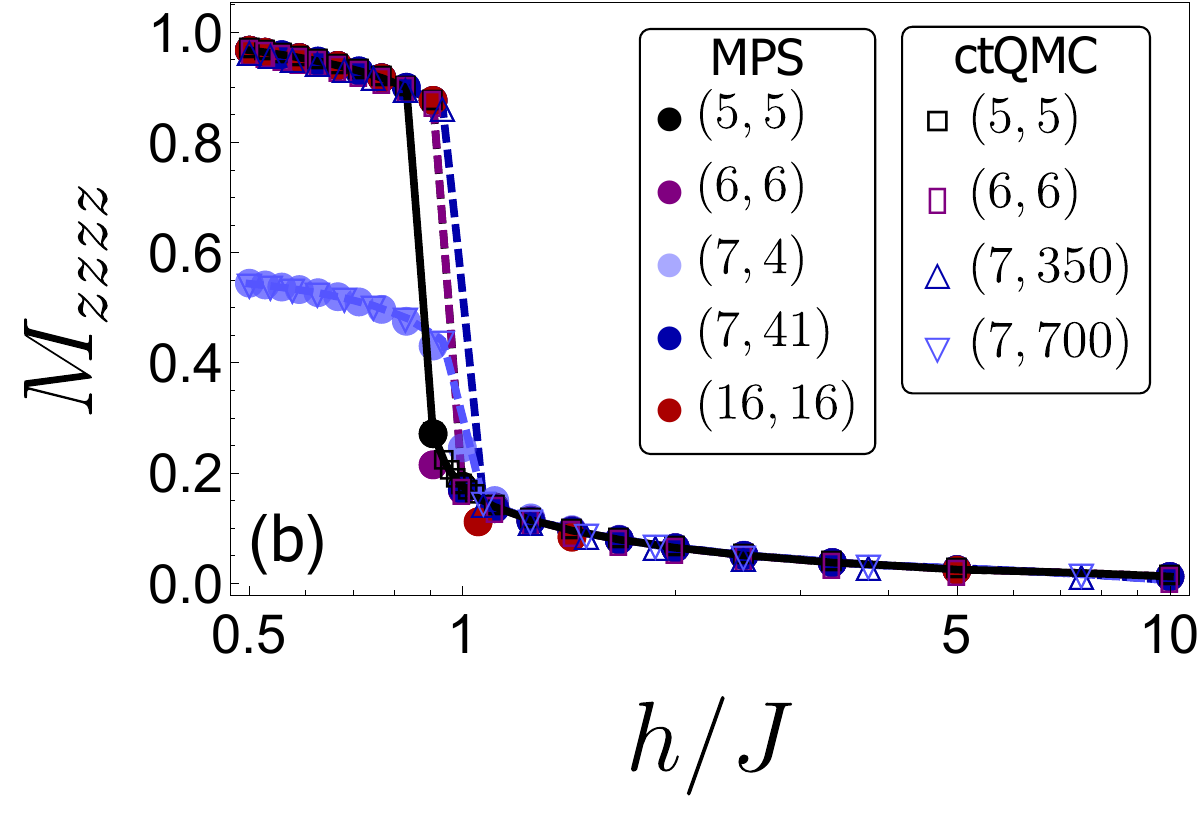}
    \end{subfigure}
    \begin{subfigure}[b]{0.24\textwidth}   
       \includegraphics[width=\textwidth]{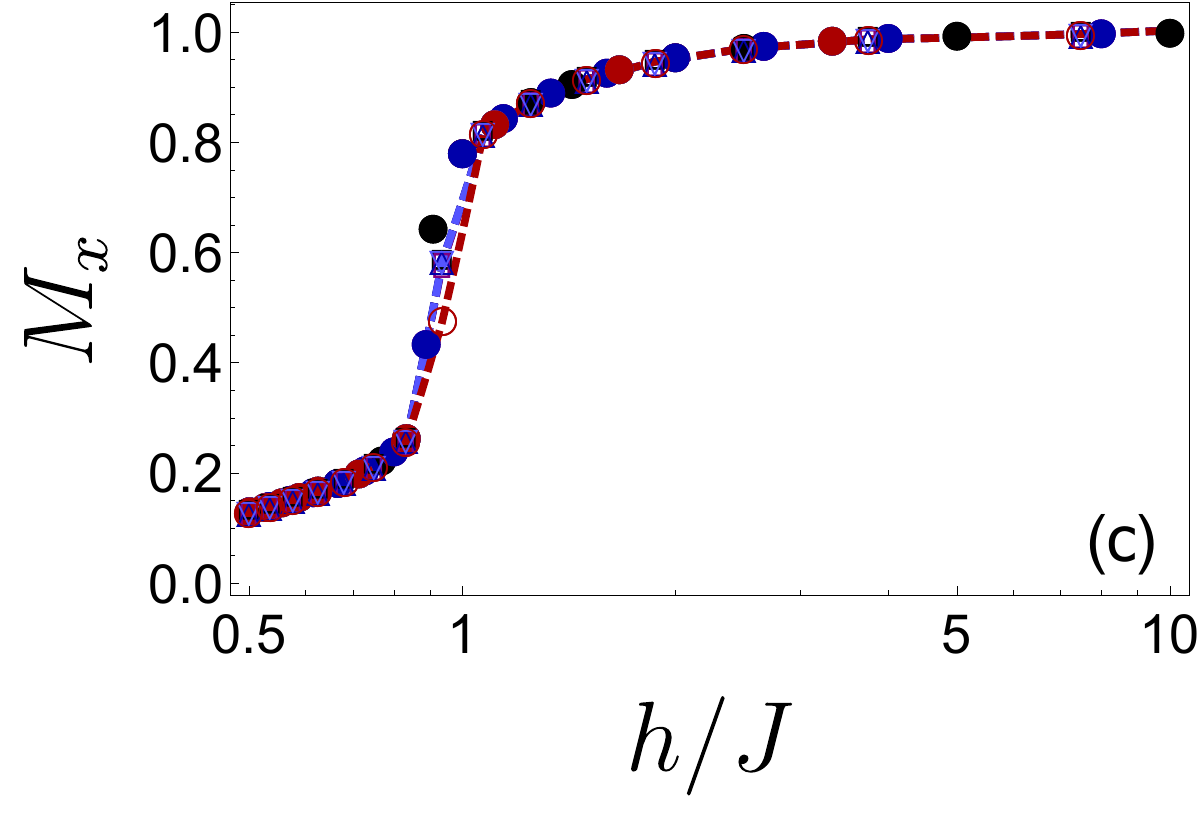}
    \end{subfigure}
    \begin{subfigure}[b]{0.24\textwidth}   
       \includegraphics[width=\textwidth]{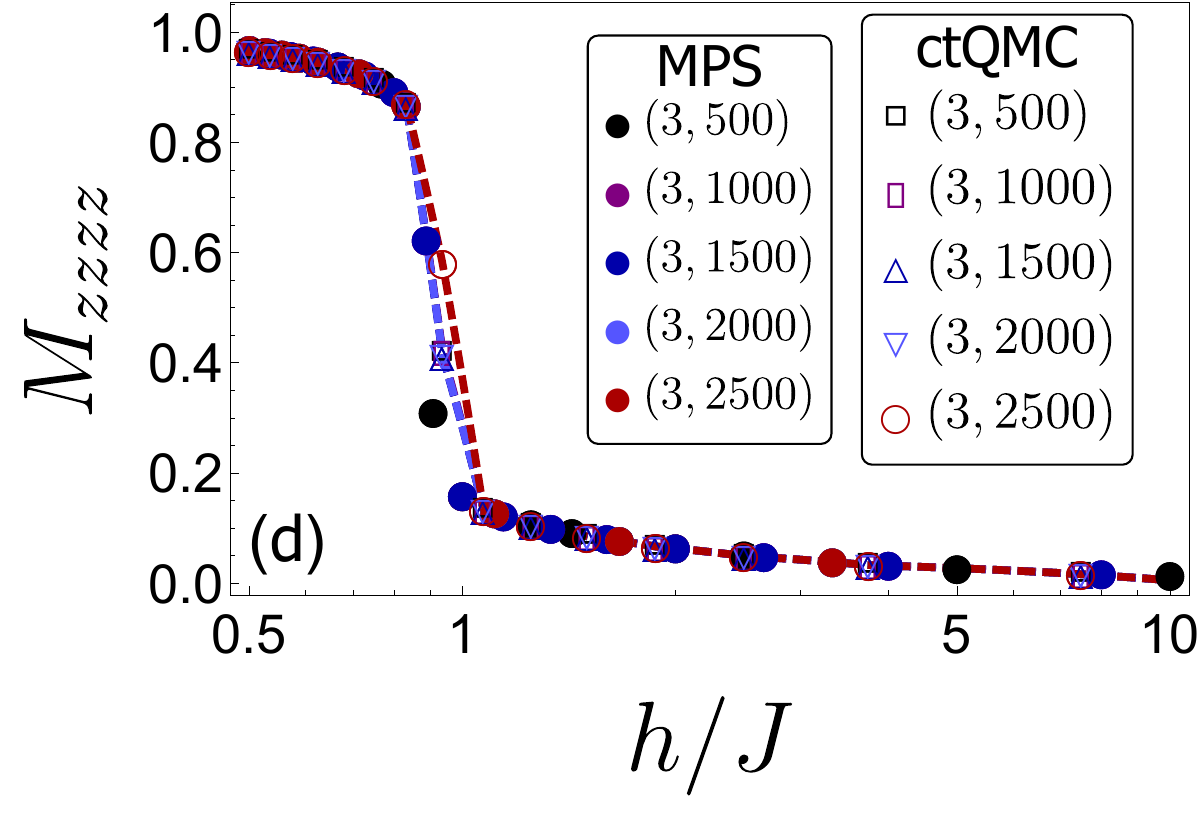}
    \end{subfigure}
    \caption{
       {\bf Quantum phase transition of $H_{54}$ for PBC, similar to Fig.~\ref{fig:quantum rule 30 PBC}.} Exact diagonalization data in (a) and (b) are shown as a continuous (black) line.
       }
     \label{fig:quantum rule 54 PBC}
 \end{figure*}

MPS simulations have been a standard numerical technique to acquire the low-energy spectrum of one-dimensional Hamiltonians, see for example Refs.~\cite{schollwock2011the-density-matrix}. They have also been applied to two-dimensional systems with success \cite{stoudenmire2012studying}. One way to do so is by using MPS in a ``snake'' form, although simulating a system of width \(L\) then requires interactions of range \(L\) to be included, and they are limited by the entanglement that they can incorporate. To exploit the expressivity and efficiency of MPS we mainly use thin-strip geometries which are quasi one-dimensional. For the continuous-time QMC algorithm we follow Ref.~\cite{causer2024rejection-free}.

For a small transverse field \(h\) we expect to verify our conclusions from the previous section: quantum fluctuations induced by \(h\) select a particular ordering pattern from the classical ground states. We refer to this as the ``classical phase''. For large enough transverse fields, we expect all three models to be in a quantum paramagnetic phase, where the spins are aligned with the transverse field and there is no symmetry breaking. Our goal is to resolve the intermediate part of their phase diagram and the presence (if any) of quantum phase transitions.

These phase transitions are expected to be of first-order, similar to the transitions into the quantum paramagnet in other plaquette spin models; see for example Refs.~\cite{2005_Nussinov,2009_Orus} for the plaquette Ising model, Ref.~\cite{sfairopoulos2023boundary} for the quantum triangular plaquette model and Ref.~\cite{sfairopoulos2023cellular} for the quantum Fibonacci and other models. For the nonlinear models treated here, we expect a transition with the addition of spontaneous symmetry breaking (SSB) for the cases where multiple classical ground states are encountered in the thermodynamic limit (cf.\ Ref.~\cite{sfairopoulos2023cellular}), specifically TSSB.

\subsubsection{Expectation values}
 
  \begin{figure*}[t]
    \begin{subfigure}[b]{0.24\textwidth}
       \includegraphics[width=\textwidth]{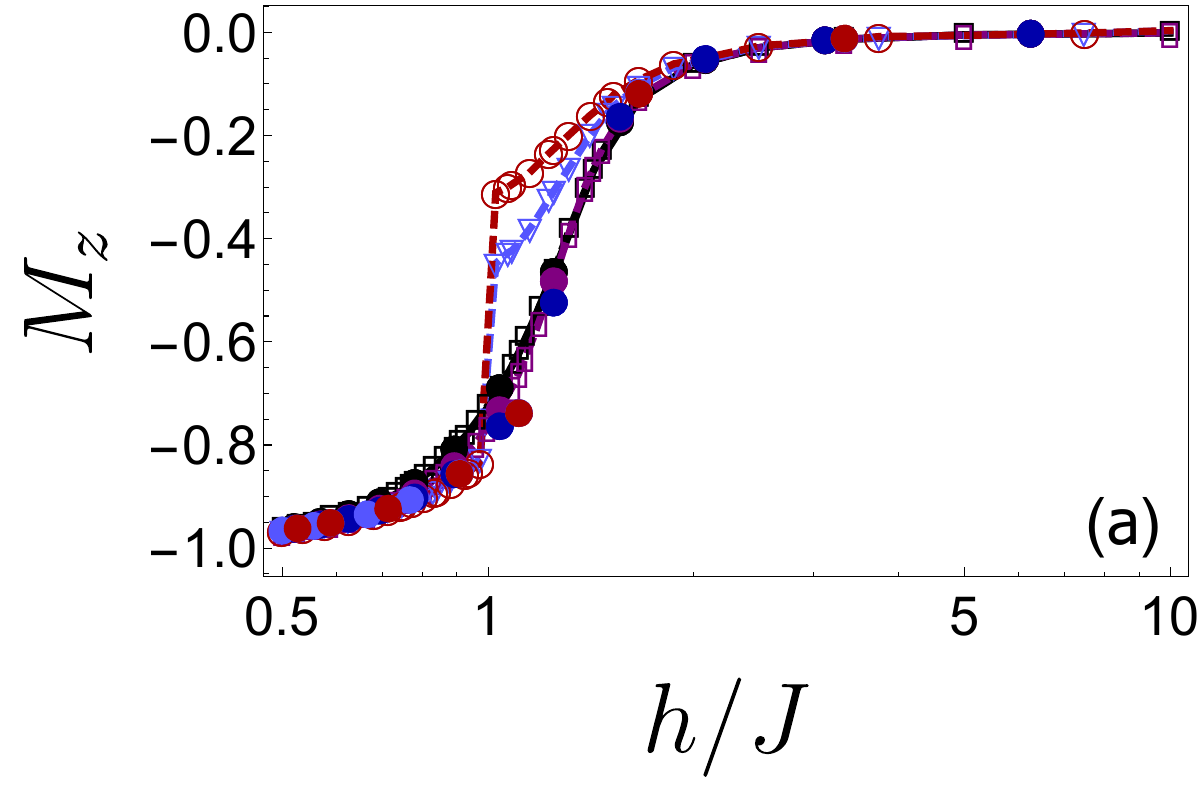}
    \end{subfigure}
    \begin{subfigure}[b]{0.24\textwidth}
       \includegraphics[width=\textwidth]{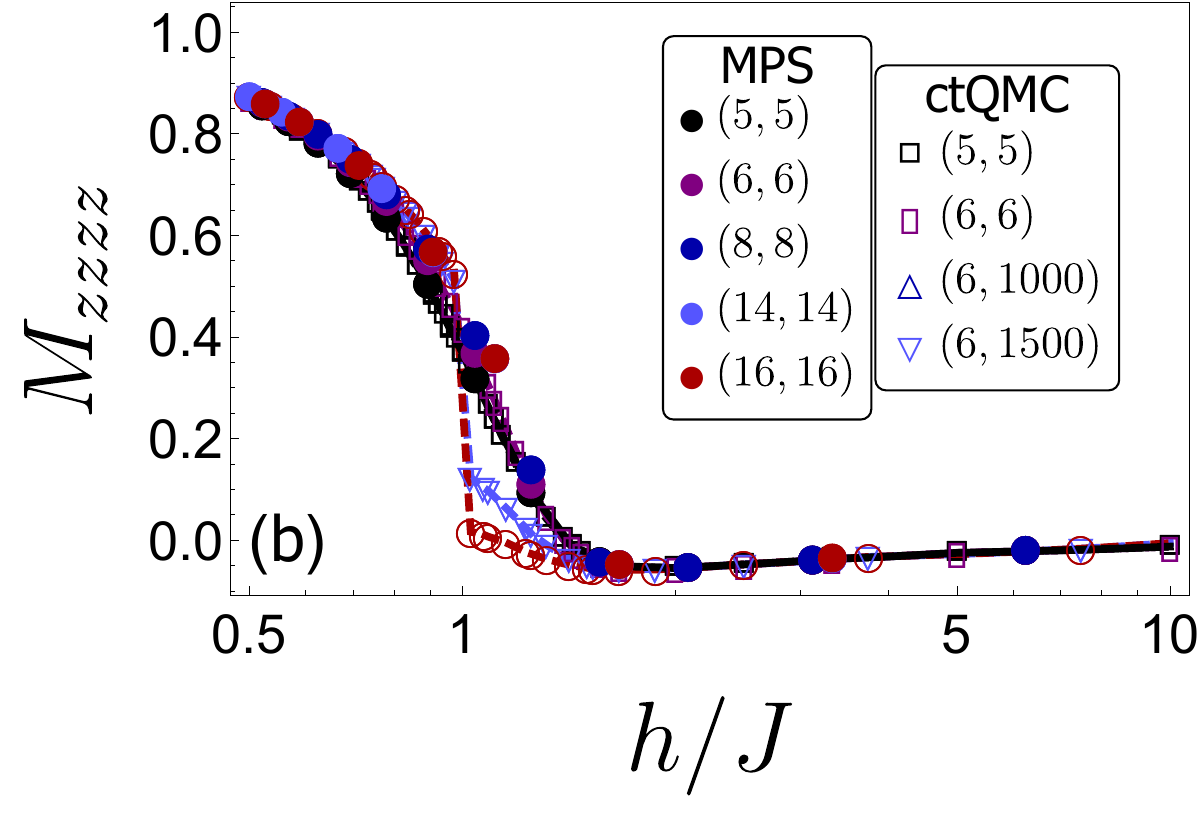}
    \end{subfigure}
    \begin{subfigure}[b]{0.24\textwidth}   
       \includegraphics[width=\textwidth]{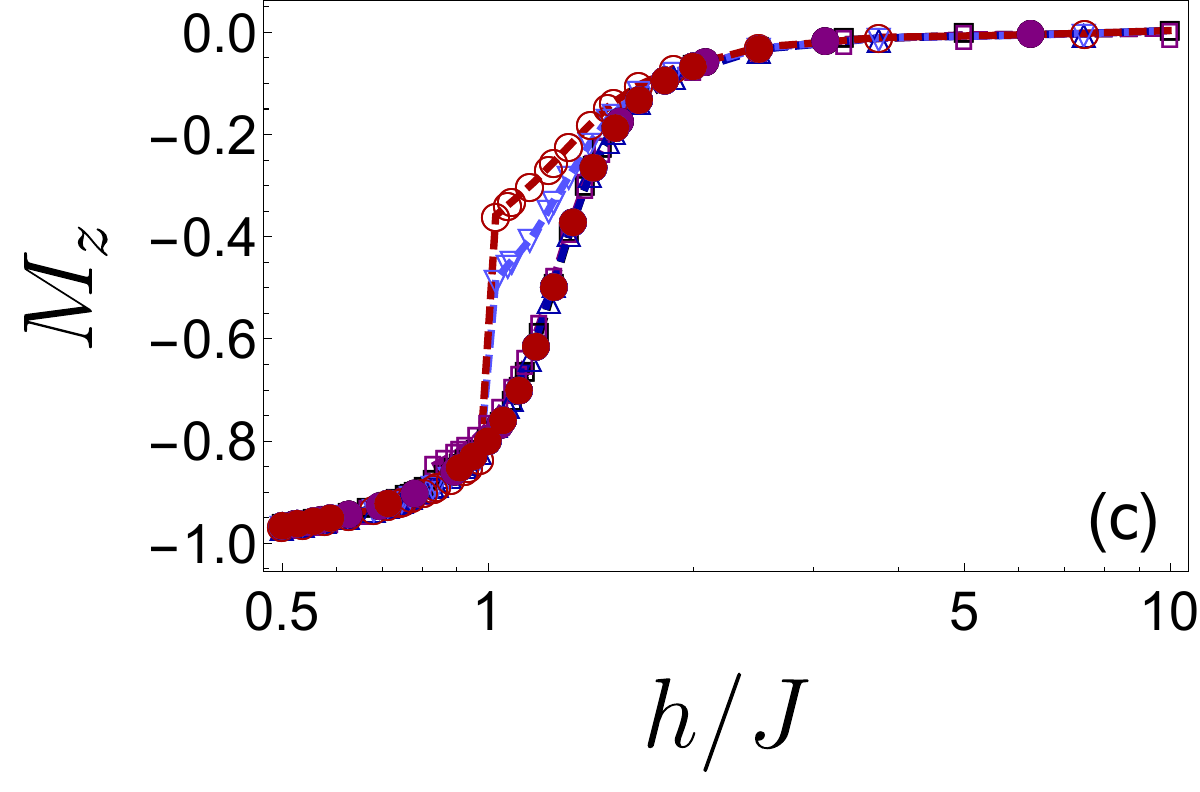}
    \end{subfigure}
    \begin{subfigure}[b]{0.24\textwidth}   
       \includegraphics[width=\textwidth]{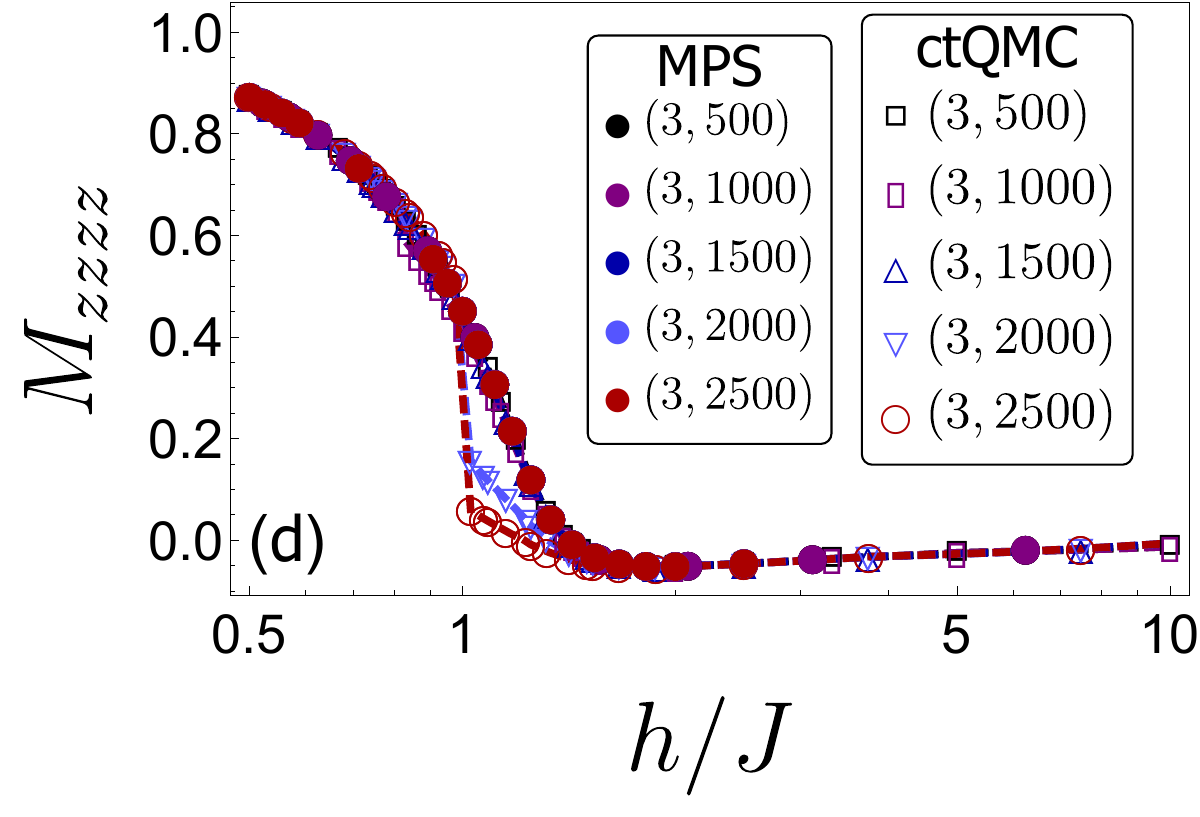}
    \end{subfigure}
    \caption{
       {\bf Quantum phase transition of $H_{201}$ for PBC, similar to Fig.~\ref{fig:quantum rule 30 PBC}.} Panels (a) and (c) show the longitudinal magnetization, $M_z$. Exact diagonalization data as in Fig.~\ref{fig:quantum rule 54 PBC}.
       }
     \label{fig:quantum rule 201 PBC}
 \end{figure*}

Figure~\ref{fig:quantum rule 30 PBC} shows our results for the quantum Rule 30. Panels (a) and (c) show the transverse magnetization per site, $M_x = \frac{1}{N} \sum_r X_r$, as a function of $h/J$ for square systems $L \times L$ and PBC from ED, from numerical MPS, and square and rectangular system sizes for ctQMC simulations and for rectangular strip geometries, respectively. Panels (b) and (d) show the four-spin correlator, $\opcatMzzzz{M} = - \sum_{\{p, q, r, s\} \in \tetrapleuro} Z_p Z_q Z_r Z_s$ for square and rectangular systems from numerical MPS and ctQMC simulations, respectively. We choose \(M_x\) and \(\opcatMzzzz{M}\) because they give clear qualitative indications of the quantum paramagnet and classical phase, respectively, and because they are largely insensitive to the precise ordering pattern selected by quantum fluctuations.

For both observables, a sudden change is observed at $h/J \approx 1.0$ which becomes steeper as the system size increases. This clearly points towards a first-order quantum phase transition in the thermodynamic limit. Indeed, these plots provide evidence that there is only one phase transition between the two limits. If this is correct, we expect the whole classical phase to be continuously connected to small-\(h/J\) limit treated in Sec.~\ref{qObD}. 

We note that for the $5 \times 5$ lattice $M_{\text{zzzz}}$ saturates at approximately $1/5$ for small \(h/J\) (in both MPS and ED), in contrast to all other system sizes shown in Fig.~\ref{fig:quantum rule 30 PBC}(b), for which it increases towards \(\approx 1\). This result can be explained on the grounds of the qObD description of the previous section. For width $L=5$, the Rule 30 CA has a single periodic trajectory of length 5, as well as a trivial fixed point with all sites unoccupied (see Appendix \ref{AppendixA}). The $5\times 5$ classical spin model therefore has 6 ground states, of which one is the all-up state and the rest are related by translations. A similar perturbation-theory calculation to Sec.~\ref{qObD30} shows that adding an infinitesimaly small transverse field selects the latter, which have $M_{\text{zzzz}} = 1/5$, in agreement with the simulation results.

This observation clearly illustrates that changes in system size can lead to unexpected changes in the behavior in the classical phase, which is a general feature of these and related quantum spin models. Nonetheless, our observations about the phase structure are unaffected: there is still clearly a first-order quantum phase transition at \(h/J\approx 1\), as evidenced by \(M_x\), even though the ordering pattern in the classical phase is substantially modified, as evidence by \(M_{\text{zzzz}}\).

Figure~\ref{fig:quantum rule 54 PBC} shows 
the same observables for the quantum Rule 54, whose behavior resembles that of Rule 30. Panels (a) and (c) show \(M_x\) while panels (b) and (d) show $M_{\text{zzzz}}$, for square and thin-strip geometries. There are again indications of a first order quantum phase transition in the thermodynamic limit, though this is less clear for the thin strips.

For this rule, there is also evidence of different structures appearing for different system sizes in the classical phase. This is apparent for system sizes \(7 \times 4p\) with $p \in \mathbb{Z}$, which display a markedly different behavior from the other sizes shown (including \(7 \times 41\) and \(7 \times 350\)), particularly for $M_{\text{zzzz}}$ but also, though much less prominently, for \(M_x\). Again, the difference can be explained by considering the periodic trajectories of the CA and using degenerate perturbation theory. For \(L=7\), there is a single nontrivial trajectory, of period \(4\), as well as an all-unoccupied fixed point (see Appendix \ref{AppendixA}). As in the calculation in Sec.~\ref{qObD54}, for this rule a small transverse field selects states that break translation symmetry. For system sizes \(7 \times 4p\), the states of period \(4\) are selected; they have $M_{\text{zzzz}} = 4/7 \approx 0.57$ which accurately reproduces the numerical calculations. For other systems with width \(L=7\), there are no such states and the only classical ground state has all spins pointing up; thus no qObD occurs and $M_{\text{zzzz}} = 1$. As noted above for Rule 30, both observables nonetheless show abrupt changes at a value \(h/J \approx 1\) for all systems.

Figure~\ref{fig:quantum rule 201 PBC} shows results for the quantum Rule 201. In this case, panels (a) and (c) show the \emph{longitudinal} magnetization, $M_z = \frac{1}{N} \sum_r Z_r$, while panels (b) and (d) show the same 4-spin correlator \(\opcatMzzzz{M}\) as in Figs.~\ref{fig:quantum rule 30 PBC} and \ref{fig:quantum rule 54 PBC}. In the limit \(h/J\rightarrow 0\), we see that $M_z \rightarrow -1$, consistent with our analysis of Sec.~\ref{qObD201}, where we argued that weak quantum fluctuations select the state \(\Ket{g^{(a)}}\) with all spins down. (For Rules 30 and 54, the selected patterns have nonuniform \(z\) magnetization, and so the scaling of \(M_z\) with system size is much less clear.) For $h/J \approx 1$ we see the formation of a crossover which, for the larger thin-strip systems, shows clear indications of the precursor of a thermodynamic phase transition. Note that we are able to simulate these system sizes only based on the ctQMC algorithm, while the finite-size scaling for smaller sizes seems to indicate a crossover rather than a phase transition.

\begin{figure*}[t]
   \centering
    \begin{subfigure}[b]{0.3\textwidth}
       \includegraphics[width=\textwidth]{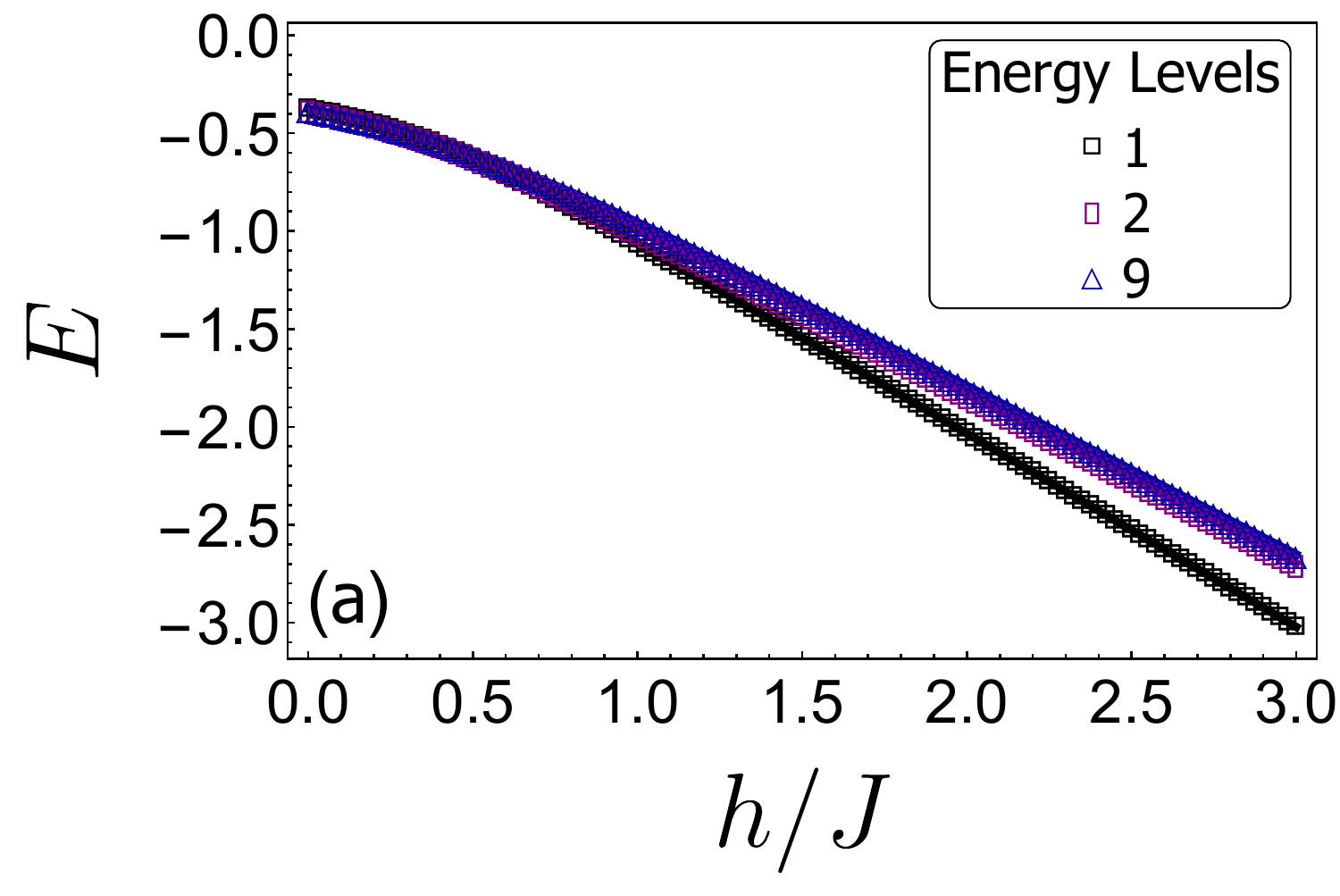}
    \end{subfigure}
    \begin{subfigure}[b]{0.3\textwidth}
       \includegraphics[width=\textwidth]{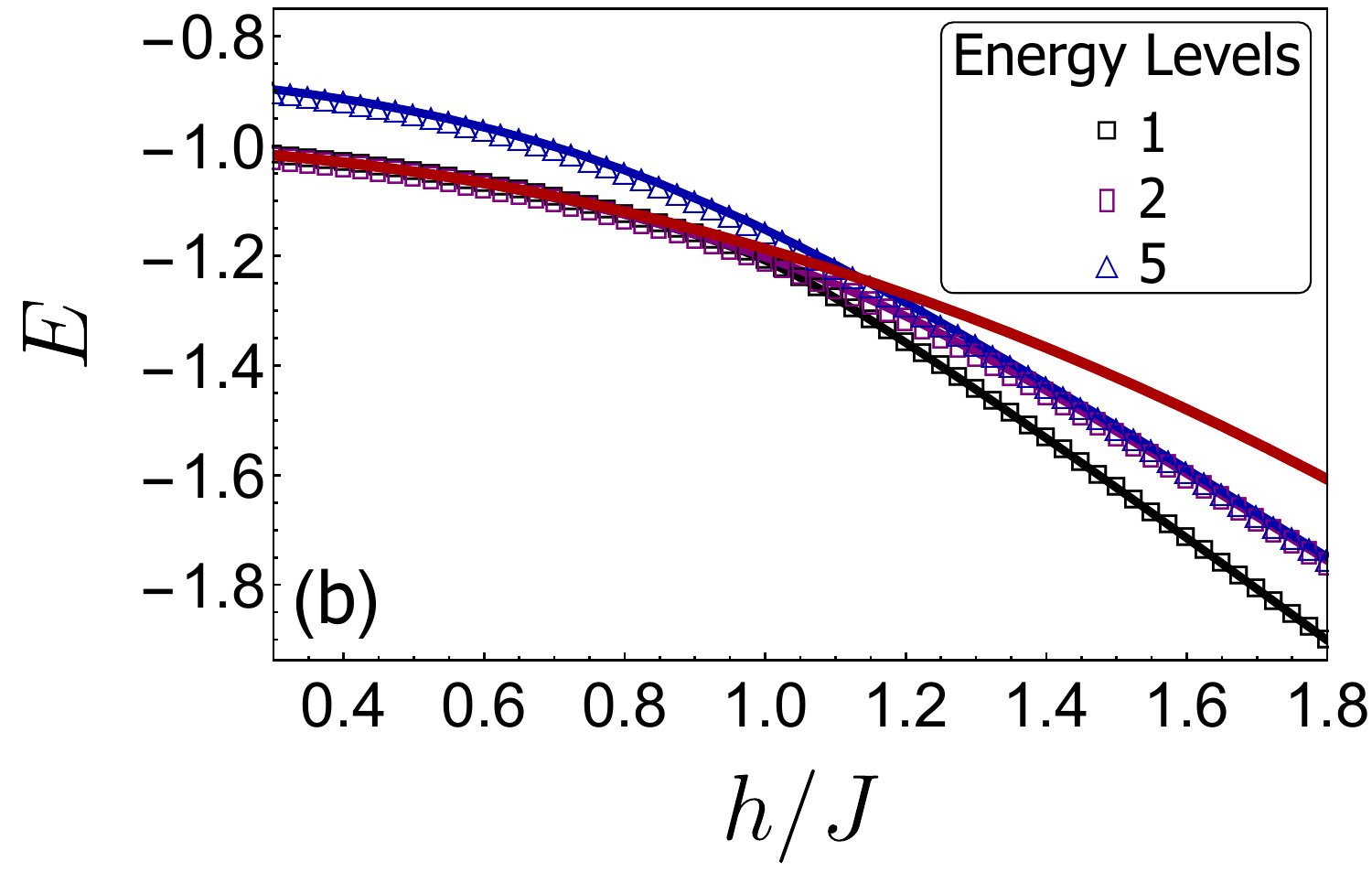}
    \end{subfigure}
    \begin{subfigure}[b]{0.3\textwidth}
       \includegraphics[width=\textwidth]{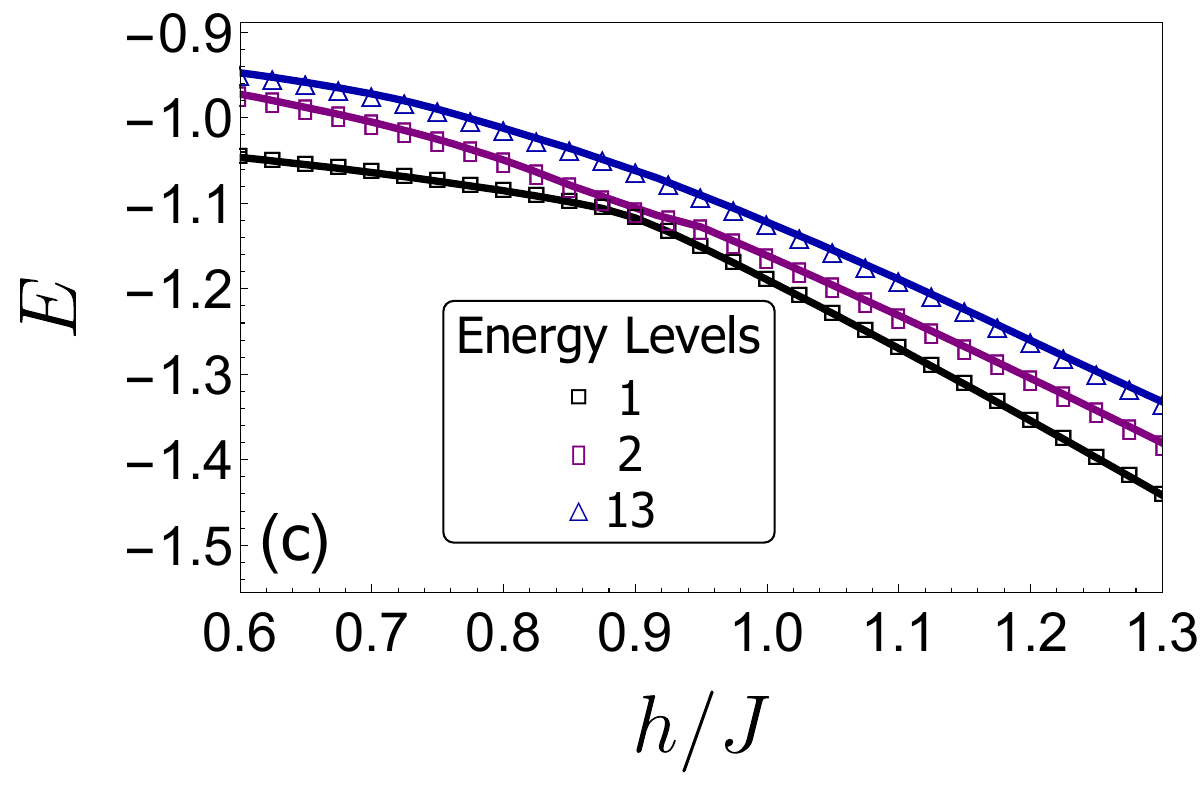}
    \end{subfigure}
    \caption{
       {\textbf{Low-lying spectrum of $H_{30}$} for a $4\times 4$ system size with OBC (a), a $4 \times 4$ (b) and a $3\times 6$ (c) system size for PBC for their respective modes. The avoided gap crossing is evident in panels (b) and (c). For panel (b) the solid dark red line indicates the prediction of second order degenerate perturbation theory for the ground state ($E_1$).}   
    }
    \label{fig:statediagram30}
\end{figure*}

\begin{figure*}[t]
   \centering
    \begin{subfigure}[b]{0.3\textwidth}
       \includegraphics[width=\textwidth]{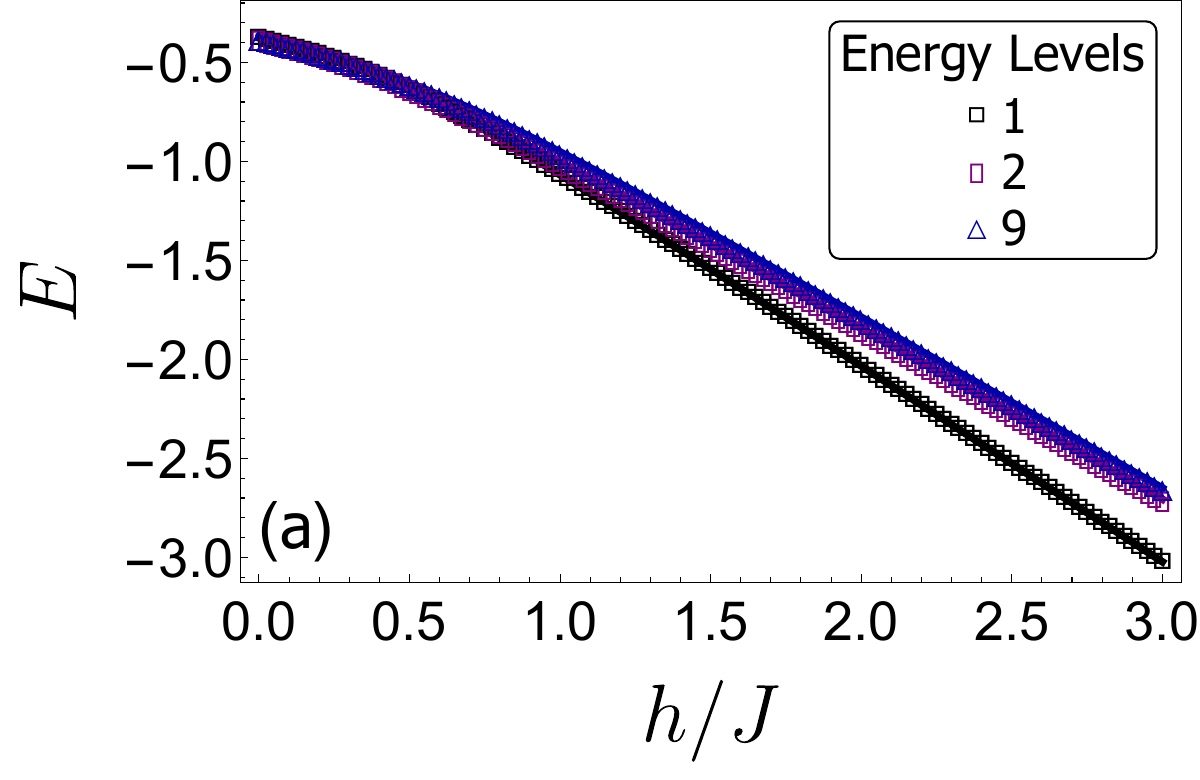}
    \end{subfigure}
    \begin{subfigure}[b]{0.3\textwidth}
       \includegraphics[width=\textwidth]{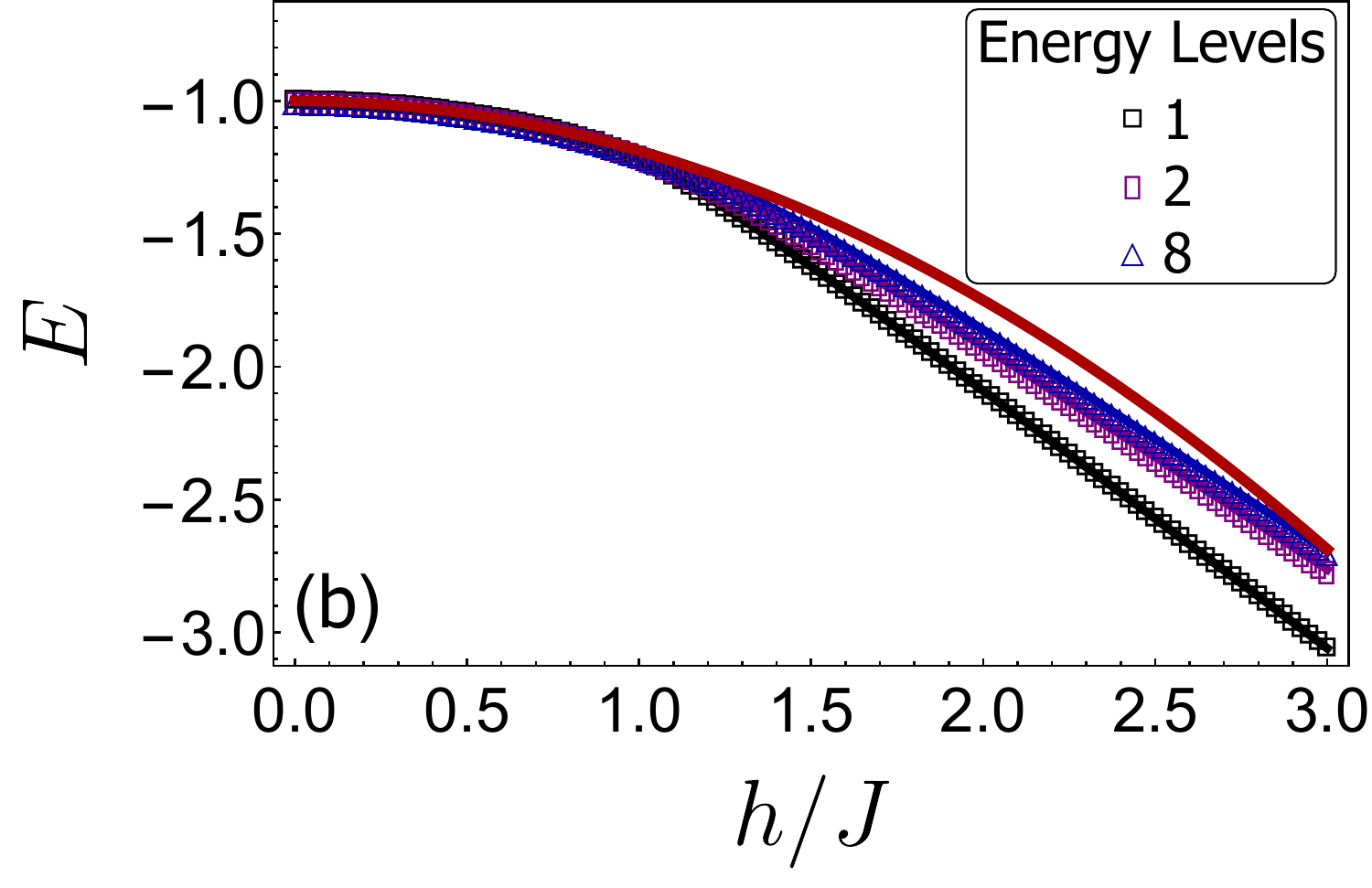}
    \end{subfigure}
    \begin{subfigure}[b]{0.3\textwidth}
       \includegraphics[width=\textwidth]{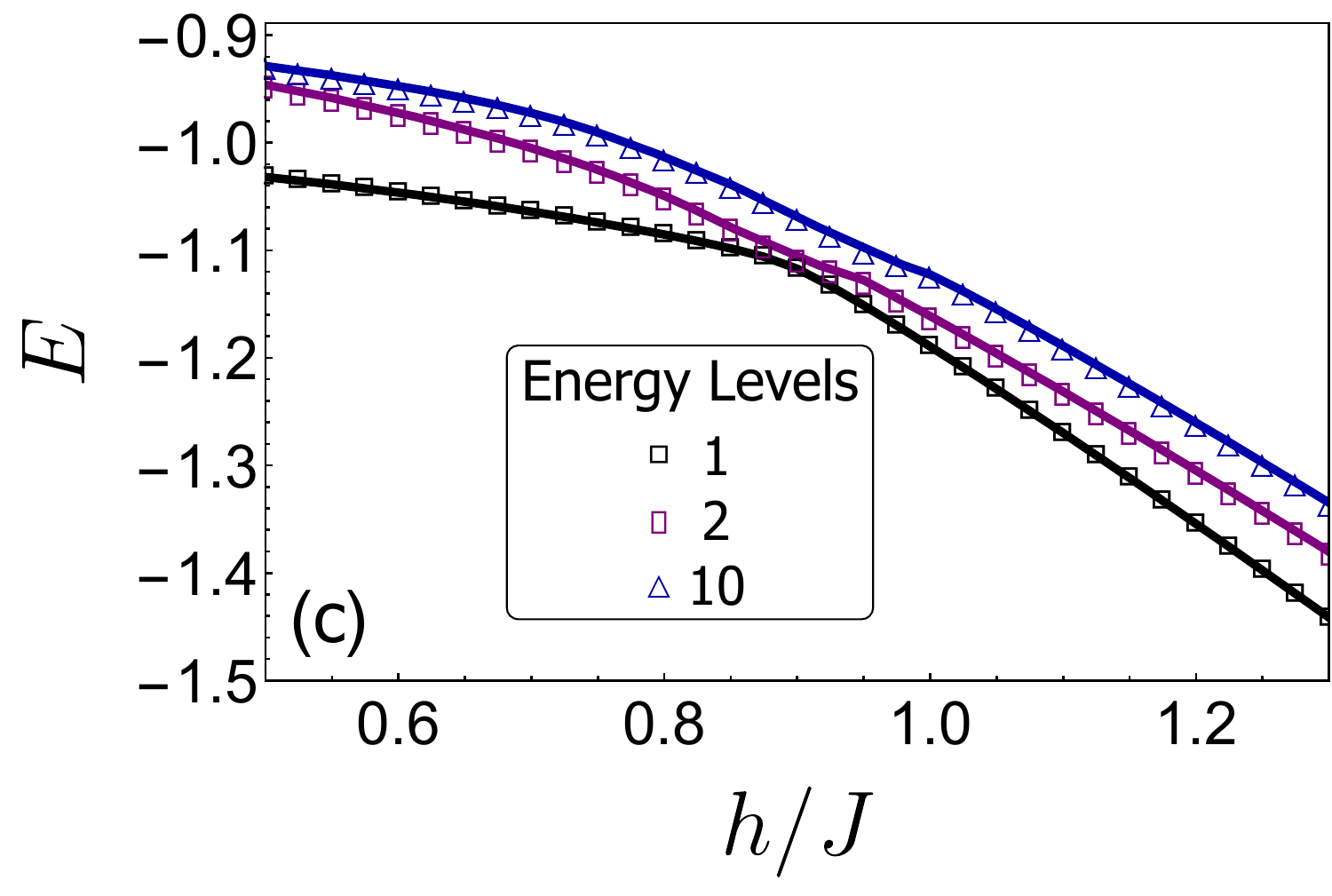}
    \end{subfigure}
    \caption{
      {\textbf{Low-lying spectrum of $H_{54}$}, similar to Fig.~\ref{fig:statediagram30}. The avoided gap crossing is evident only in panel (c). Similar in panel (b) the prediction of perturbation theory for the ground state included.}    
    }
    \label{fig:statediagram54}
\end{figure*}

\begin{figure*}[t]
   \centering
    \begin{subfigure}[b]{0.3\textwidth}
       \includegraphics[width=\textwidth]{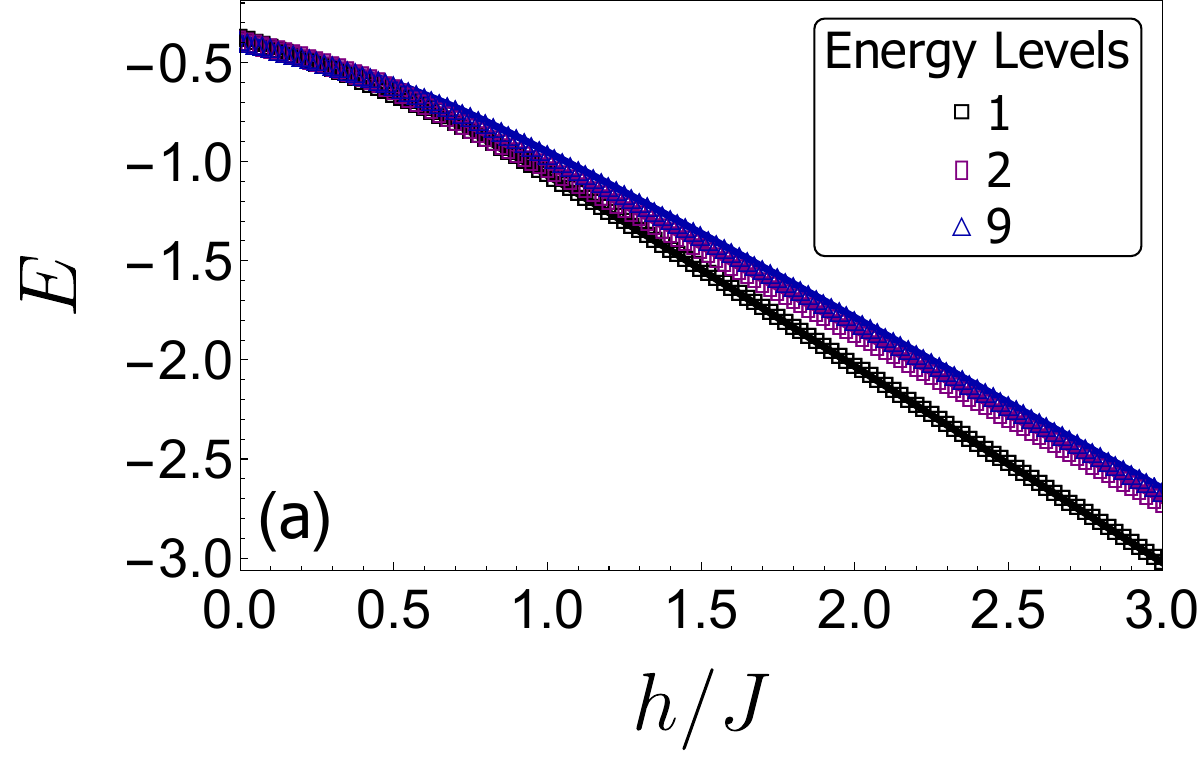}
    \end{subfigure}
    \begin{subfigure}[b]{0.3\textwidth}
       \includegraphics[width=\textwidth]{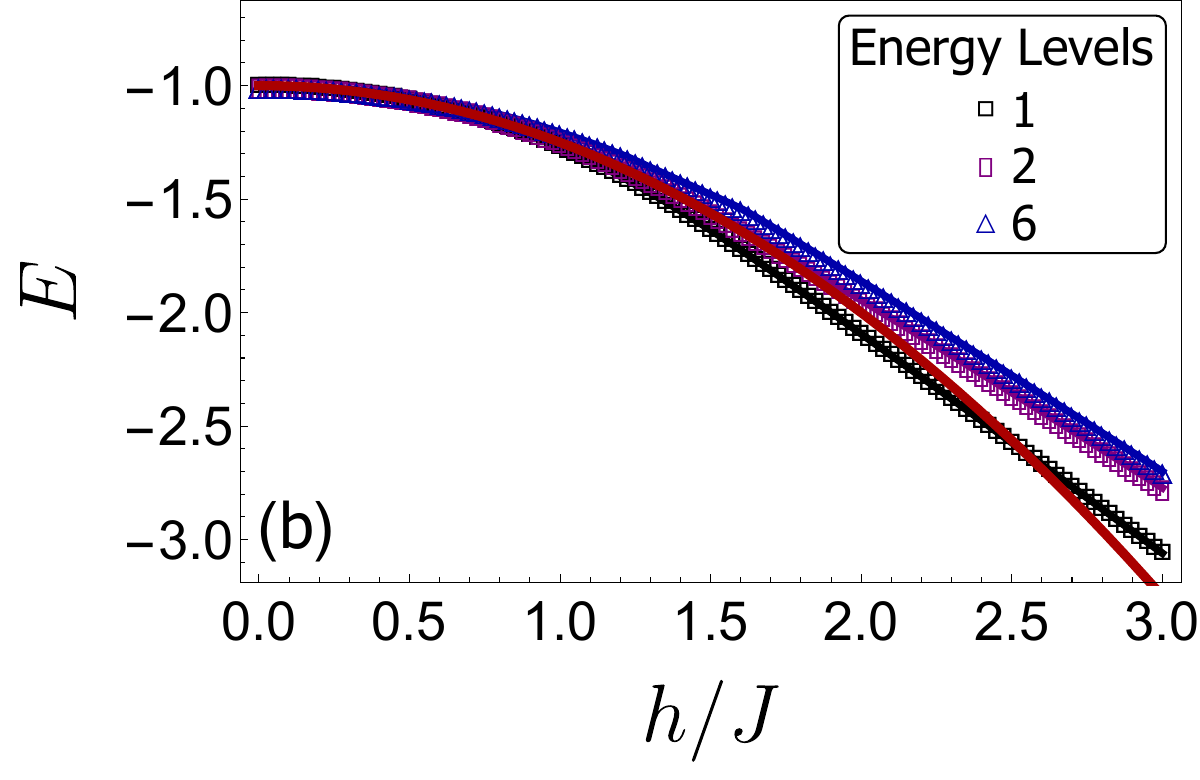}
    \end{subfigure}
    \begin{subfigure}[b]{0.3\textwidth}
       \includegraphics[width=\textwidth]{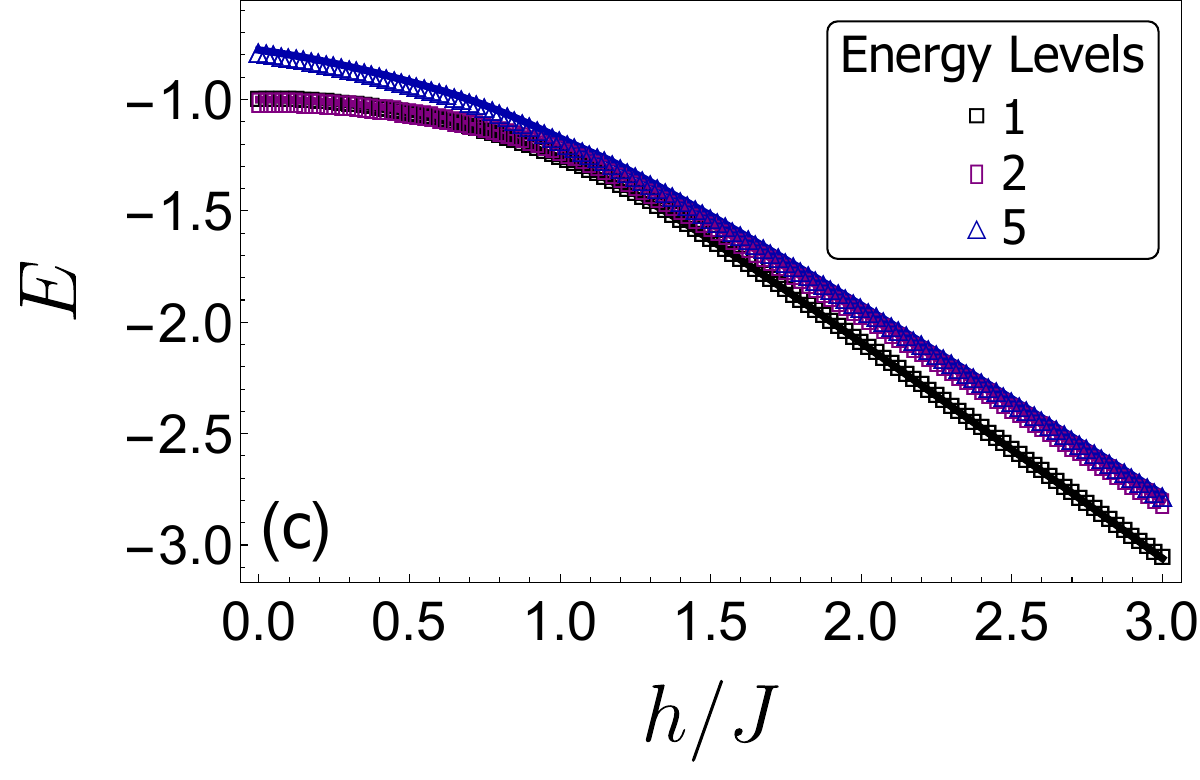}
    \end{subfigure}
    \caption{
      {\textbf{Low-lying spectrum of $H_{201}$}, similar to Fig.~\ref{fig:statediagram30}. The avoided gap crossing is evident only in panel (c).}  
    }
    \label{fig:statediagram201}
\end{figure*}
 
The presence of the qObD is apparent in all of these figures. More specifically, for the case of the quantum Rule 30 in Fig.~\ref{fig:quantum rule 30 PBC}(b), different system sizes seem to approach different values for \(\opcatMzzzz{M}\) in the classical phase. As noted above, we understand this system-size dependence as resulting from the selection of different ground states for the system sizes studied from the respective order of the degenerate perturbation theory. A similar phenomenon is observed in Fig.~\ref{fig:quantum rule 54 PBC}(b). Lastly, in all panels of Fig.~\ref{fig:quantum rule 201 PBC} we observe that the classical phase saturates the longitudinal magnetization \(M_z\) as well as the four-spin correlator. As noted above, this effect can be explained through the prism of the qObD mechanism and the selection of the symmetric, all-down ferromagnetic state.

For all models studied here, the convergence of the numerical algorithms (with the exception of exact diagonalization) is particularly challenging, and considerably harder than for the linear rules \cite{sfairopoulos2023cellular}. This is most notable for the numerical MPS, something that we attribute to the frustration of the Hamiltonians studied. These convergence issues can be seen for example in panel (b) of Fig.~\ref{fig:quantum rule 30 PBC} deep into the frustrated phase and for Rule 54 close to the phase transition point. For Rule 201, close to the $h = J$ point the numerics is not clear enough and convergence to the ground state is not obvious. The signatures of the quantum phase transition are confusing and the ctQMC method shows unexpected effects for larger strip geometries, when compared to the ED data. Further numerical evidence would be needed for the clarification of the nature of the quantum phase transition close to the $h = J$ point. We believe that the enhanced hardness for the quantum Rule 201 is due to the increased number of classical ground states which become low-lying excited states for the quantum model and possibly hinder the convergence of our numerical methods, in a similar way to the quantum triangular plaquette model for OBC \cite{sfairopoulos2023boundary}.

\subsubsection{Low-lying spectra}

Figures~\ref{fig:statediagram30}--\ref{fig:statediagram201} show the low-lying spectra of these models for small system sizes calculated via ED. Our goal in these figures is to show signatures of the formation of an avoided crossing, a characteristic of first-order quantum phase transitions \cite{sachdev2011quantum}. At the same time, we want to show the classical ground state degeneracy (if any) for the system sizes studied and the respective symmetry breaking, which we connect to the results of Sec.~\ref{qObD}. In each case, we show the lowest two levels and one higher level selected to illustrate these points.

Figure~\ref{fig:statediagram30} shows the low-lying spectrum for two small system sizes for $H_{30}$. In panel 
(b) we show the case of the size $4 \times 4$ for PBC; here we show the first, second and fifth lowest energy configurations. Quantum effects for this rule, discussed in Sec.~\ref{qObD30}, tend to select configurations with vertical stripes (the same ordering pattern as in panel (b) of Fig.~\ref{fig:rule30_tiles}). In the classical phase for small \(h/J\), this results in two degenerate ground states, as visible in the figure, related by translation symmetry \footnote{Note that for finite systems this degeneracy is split by corrections that are exponentially small in system size, but these are not visible in the figure.}. Their energy matches closely with the result of second-order perturbation theory, shown in solid red. On the other hand, for the paramagnetic phase at large \(h/J\), the degeneracy is always lifted and a unique ground state is seen. Close to $h/J = 1.0$ we find the finite-size signature of the eventual first-order quantum phase transition, namely a relatively sudden change in the gradient, which can be identified as an avoided level crossing, accompanied by splitting of the degeneracy.

In panel (c) we show the same low lying spectrum for a system of size $3 \times 6$. In this case, there is a single classical ground state, as predicted for odd \(L\), and the precursor of the avoided gap crossing is found between the first and the second energy levels. Panel (a) shows the energy spectrum for the lowest energy states for OBC. Similar behavior to the PBC is observed, the difference being that the classical ground state space includes many more states \footnote{Specifically, $2^{L + 2M - 2}$ for all three models studied here.} and results in hiding the avoided level crossing behind a subextensively large low-energy spectrum. Lastly, note that for OBC there is no ObD mechanism lifting the ground state degeneracy.

For completeness, similar results (with the same system sizes and boundary conditions) are presented for Rules 54 and 201 in  Figs.~\ref{fig:statediagram54} and \ref{fig:statediagram201}, respectively, although avoided crossings are harder to identify in these cases than for Rule 30. Only for Rule 54 with system size $3\times 6$, Fig.~\ref{fig:statediagram54}(c), can one easily be observed. When compared to each other, Rule 201 has a larger classical ground state manifold, thus explaining the increased difficulty of identifying an avoided crossing for this model. This is due to the many more classical ground states which become low lying states for nonzero transverse field terms, and also give rise to the increased complexity of the previous simulations for larger system sizes. In both panels (b) in Figs.~\ref{fig:statediagram54} and \ref{fig:statediagram201}, we have also included the ground-state energy calculated using second order degenerate perturbation theory, which is in good agreement for \(h/J \lesssim 1.0\).
\section{Conclusions}{\label{conclusions}}

In this work we have defined and studied a class of spin models whose zero temperature classical ground states are derived from the periodic orbits of elementary CA with nonlinear rules. Making these models quantum by adding a transverse field \(h\), we studied their ground state phase diagrams using perturbation theory for small \(h\) and then ED, MPS and ctQMC simulations for the whole phase diagram.

We have demonstrated the existence of an ObD mechanism for small \(h\) where quantum fluctuations split the classical ground state degeneracy. For some cases, such as the spin model from Rule 201, we can argue that a particular ordering pattern will be selected in the thermodynamic limit, while for others, such as Rule 30, a similar argumentation failed. A more complete treatment of the perturbative effects of the quantum fluctuations in these models would require the use of linked cluster expansion methods or continuous unitary transformations \cite{oitmaa2006series,adelhardt2024monte}.

For larger \(h\), we showed evidence for a first-order quantum phase transition in the thermodynamic limit, as in other spin models derived from CA \cite{sfairopoulos2023boundary,sfairopoulos2023cellular}. The transitions for the models studied here can be divided into two categories. One follows the paradigm of the quantum Rule 201 where the qObD leads to a trivial classical phase and so the accompanying phase transition exhibits no symmetry breaking. This case resembles the one of the quantum triangular plaquette model for system sizes a power of 2 \cite{vasiloiu2020trajectory,sfairopoulos2023boundary}. Models in the other category, including the quantum Rule 54 (and potentially Rule 30), show TSSB at the first-order phase transition. The two classes of models are considerably different: the first one possesses a phase transition with no symmetry change (cf.\ liquid-gas), while the second with a change (cf.\ solid-liquid).

Comparing our results here with studies of quantum spin models derived from linear CA  \cite{sfairopoulos2023cellular}, we note some similarities as well as significant differences. For both types of rules, there is clear evidence for a first-order quantum phase transition between a classical phase at small \(h\) and a quantum paramagnet at large \(h\) \footnote{A few of the linear rules give spin models that are decoupled Ising chains, in which case the transition is continuous.}. Applying perturbation theory starting from the ground states of the classical model gives an effective Hamiltonian \(H_{\text{eff}}\) in which off-diagonal elements appear only at an order that increases at least with linear system size.

For the linear rules, however, there is an additional symmetry that relates all classical ground states to each other \cite{sfairopoulos2023boundary}, and means that the diagonal elements of \(H_{\text{eff}}\) are all equal. For the nonlinear rules, there is no such symmetry, and so \(H_{\text{eff}}\) splits the classical ground states, picking out a particular spatial pattern by qObD, with TSSB in some cases, as noted above. The absence of the additional symmetries in the case of nonlinear rules can be attributed to the presence of CZ gates in their interactions (see Section~\ref{classical_models}).

The essential ingredients in our conclusions are the fact that ground states differ by a number of spin flips that scales at least linearly with the vertical dimension of the system, which follows from the directionality inherent in the deterministic CA (see Sec.~\ref{DegeneratePerturbationTheory}), and the presence of energy differences between different excited states, due to the nonlinearity of the rules. We therefore expect other nonlinear CA to give quantum spin models with similar phase diagrams, including first-order transitions and, in some cases, qObD in the classical phase. Other models, such as Ref.~\cite{stephen2024ergodicity,sfairopoulos2024the-quantum}, with nonlinear local constraints but no directionality, can in principle exhibit an extensive ground state manifold and also be candidates for spin liquid states \cite{narasimhan2024simulating}. Similarly, the constraint can be of longer support, where we would expect a  similar behavior. 

As with the linear rules, the sensitivity of the number and structure of the classical ground states on system size makes finite-size scaling challenging, and different scaling procedures can give different results for some models \cite{sfairopoulos2023boundary}. While periodic orbits of the linear rules can be obtained by gaussian elimination, no well-established number-theoretic argument applies to the nonlinear rules, and so we are unable to determine a general pattern for their compatible system sizes. The only general conclusion concerns Rule 201, where we conjecture (without rigorous proof, but confirmed for sizes up to \(L = 25\)) that the only periodic orbits encountered are of length 1 and 2, and that their multiplicity increases faster than linearly but always subextensively. For Rule 30 and Rule 54, the increase of the number of classical ground states is in general nonmonotonic (see Appendix~\ref{AppendixA}). We have therefore not been able to characterize the quantum phase transitions in detail, including any accompanying SSB, for all the models studied. Instead, we have focused on demonstrating the existence of transitions, their first-order nature, and, in certain cases, the phenomenon of qObD. We are also able to argue for the presence of TSSB for some models, such as the quantum Rule 54, and its absence for others, such as Rule 201.

In this work we focused on the zero-temperature classical properties of these models and then studied them in the presence of quantum fluctuations. However, we have not studied the effects of thermal fluctuations. We expect the occurrence of thObD for finite temperature, although a detailed study of the combination of both thermal and quantum fluctuations might reveal additional interesting phenomena \cite{rau2018pseudo-goldstone,khatua2023pseudo-goldstone}. We did not study the quantum dynamics of these models either, or the possibility of any nonthermal states or nonergodic behavior. 

Another future avenue concerns the potential presence of duality arguments for the nonlinear models. Models from linear CA possess duality arguments by exchanging the two terms of their Hamiltonians (see for example \cite{cobanera2011the-bond-algebraic,lootens2023dualities,gorantla2024tensor} for general treatments), similar to the one studied extensively for the quantum Ising model \cite{kramers1941statistics,seiberg2024majorana,seiberg2024non-invertible} or the quantum triangular plaquette model \cite{vasiloiu2020trajectory}. These duality arguments provide a powerful constraint: if there is only one phase transition, then it has to be located at $J=h$. This has been verified numerically for the quantum triangular plaquette model \cite{vasiloiu2020trajectory,sfairopoulos2023boundary} but also for the quantum Rule 150 and the square pyramid model \cite{sfairopoulos2023cellular}. All indications for the presence of quantum phase transitions found in this work locate these in the neighborhood of the $J=h$ point. Thus, a logical question to ask is whether any duality arguments, and, subsequently, any emergent symmetries exist for the nonlinear models at their phase transition points, and, if so, what their properties would be.

\begin{acknowledgments} 
  KS wants to thank A. Gammon-Smith and C. Castelnovo for valuable feedback on the manuscript. We acknowledge financial support from EPSRC Grants no.\ EP/R04421X/1 and EP/T021691/1, the Leverhulme Trust Grant No. RPG-2018-181, and University of Nottingham grant no.\ FiF1/3. LC was supported by an EPSRC Doctoral prize from the University of Nottingham. Simulations were performed using the University of Nottingham Augusta and Ada HPC cluster, and the Sulis Tier 2 HPC platform hosted by the Scientific Computing Research Technology Platform at the University of Warwick (funded by EPSRC Grant EP/T022108/1 and the HPC Midlands+ consortium). 
\end{acknowledgments}

\appendix
\section{Periodic structure of CA rules 30, 54 and 201}{\label{AppendixA}}

\begin{figure*}
    \centering
    \begin{subfigure}[b]{0.24\textwidth}
        \includegraphics[width=37mm, height=37mm]{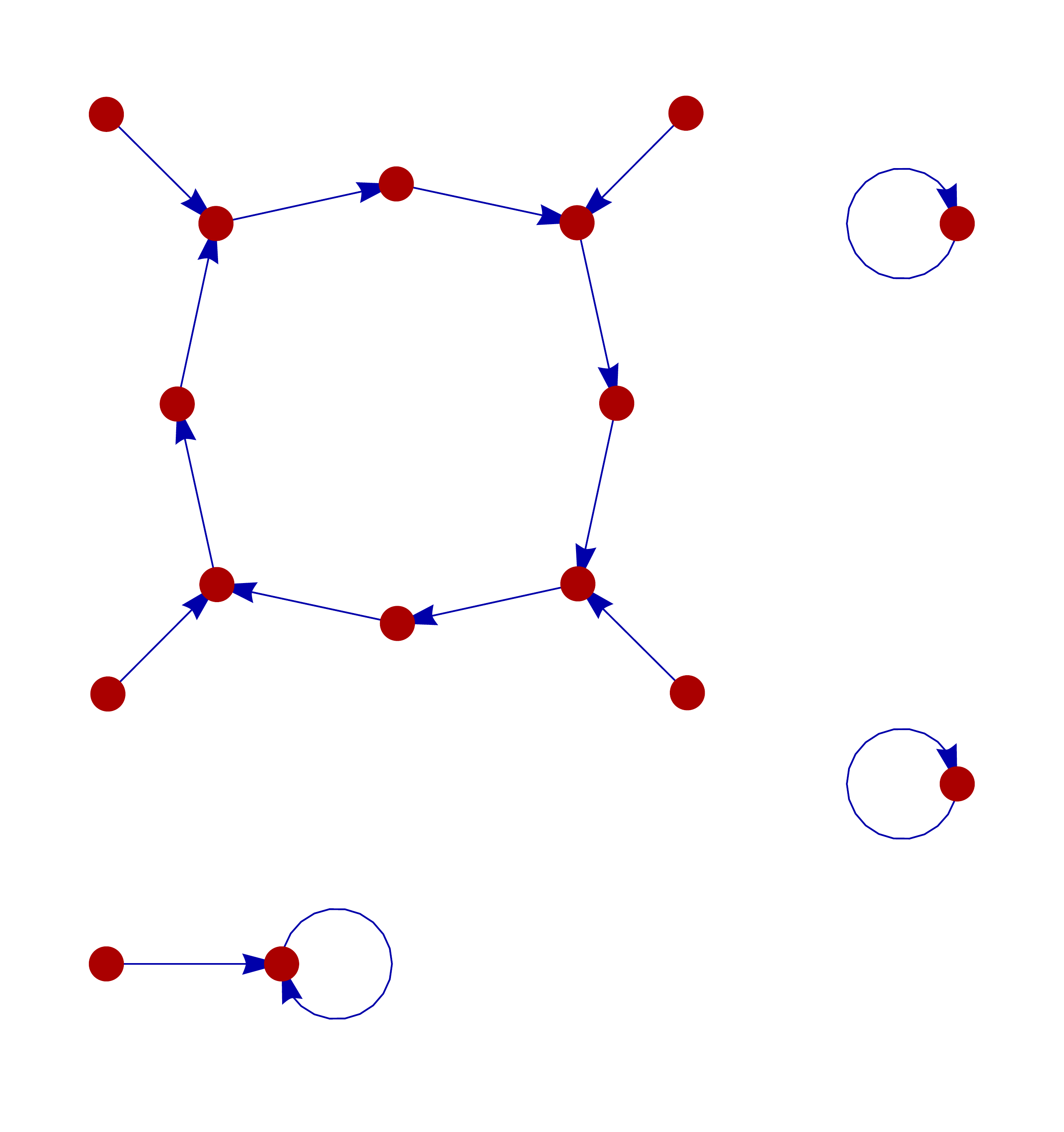}
        \caption{}
     \end{subfigure}
     \begin{subfigure}[b]{0.24\textwidth}
        \includegraphics[width=37mm, height=37mm]{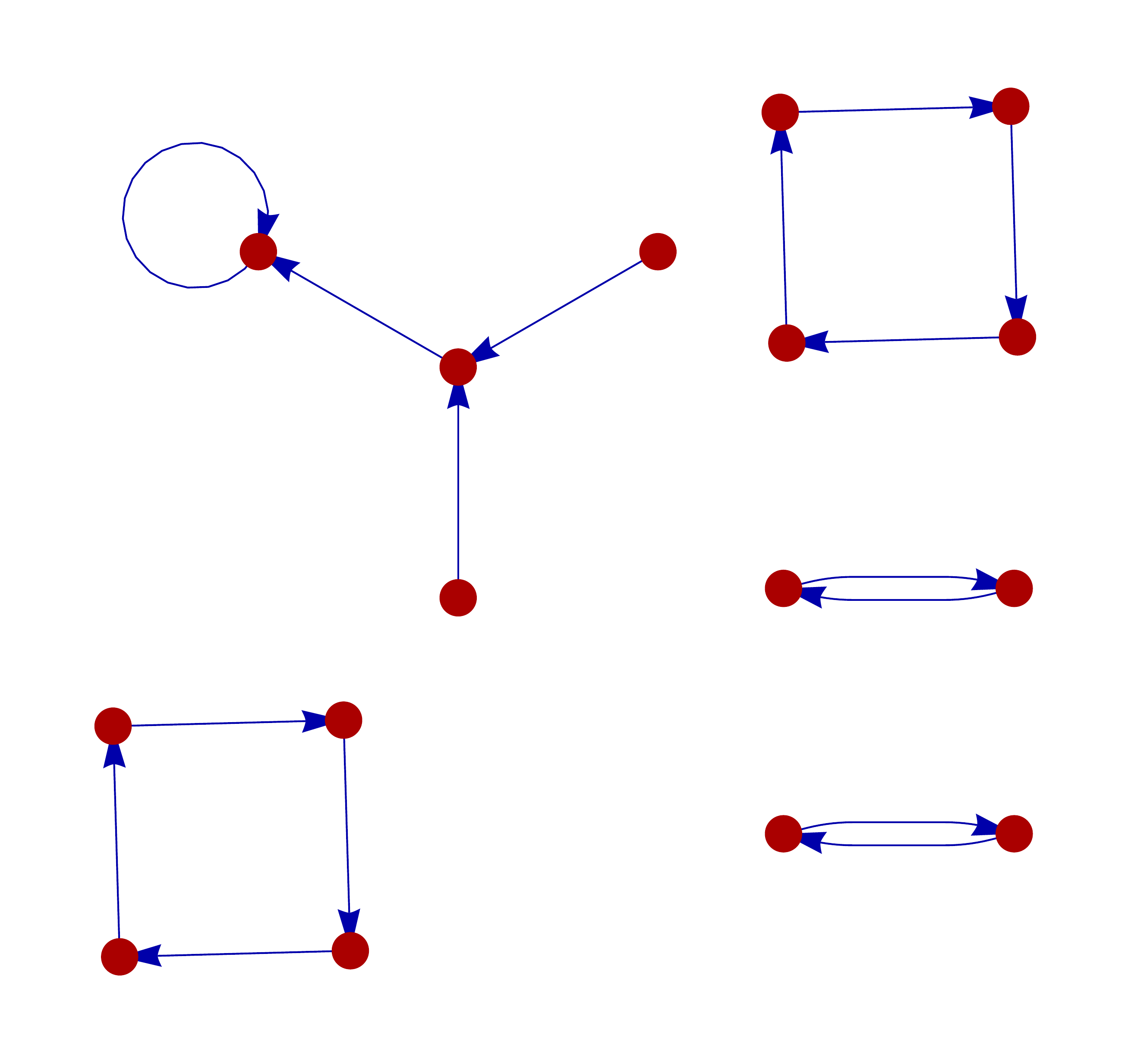}
        \caption{}
     \end{subfigure}
     \begin{subfigure}[b]{0.24\textwidth}
         \includegraphics[width=37mm, height=37mm]{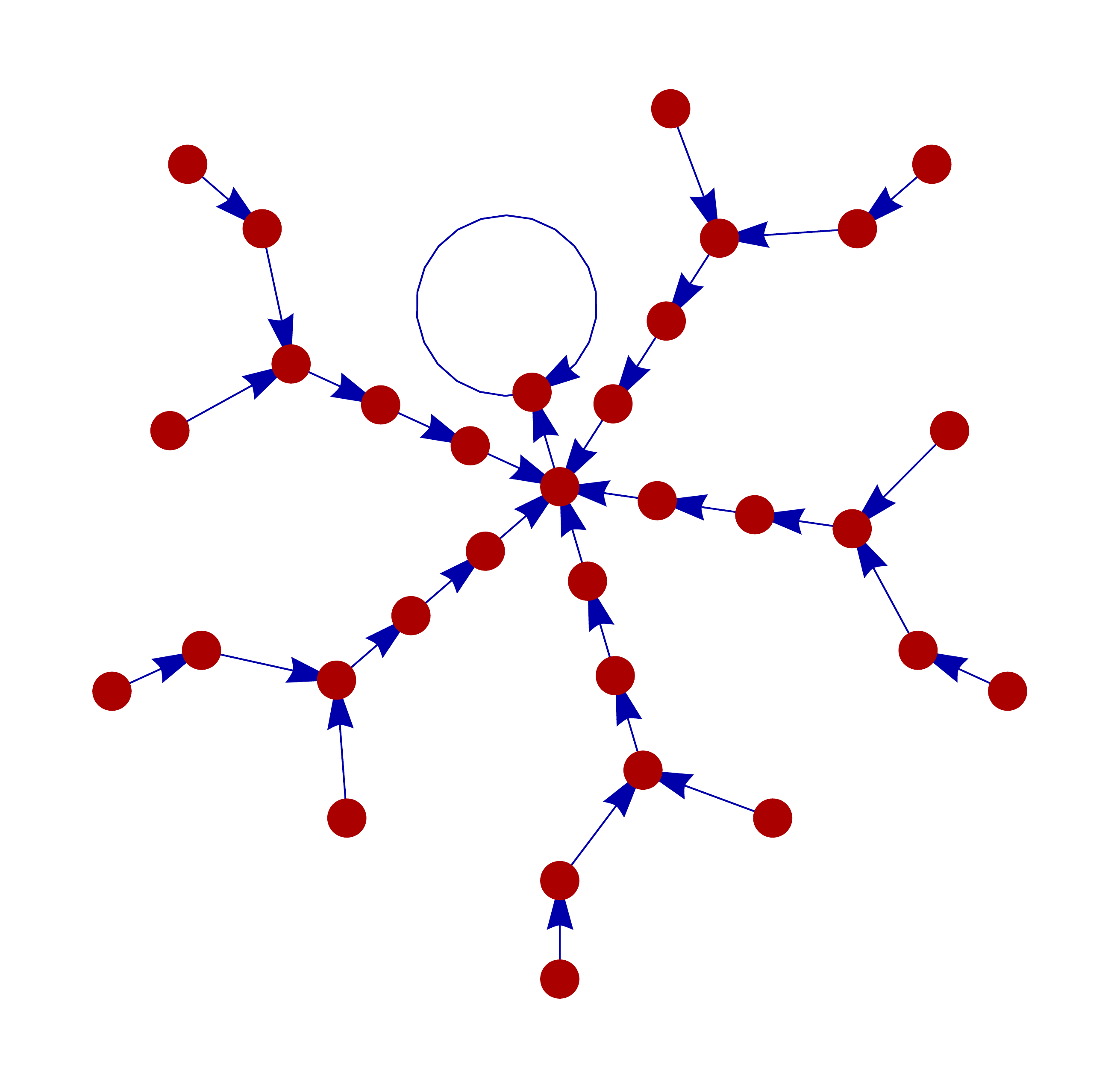}
         \caption{}
      \end{subfigure}
     \begin{subfigure}[b]{0.24\textwidth}
        \includegraphics[width=37mm, height=37mm]{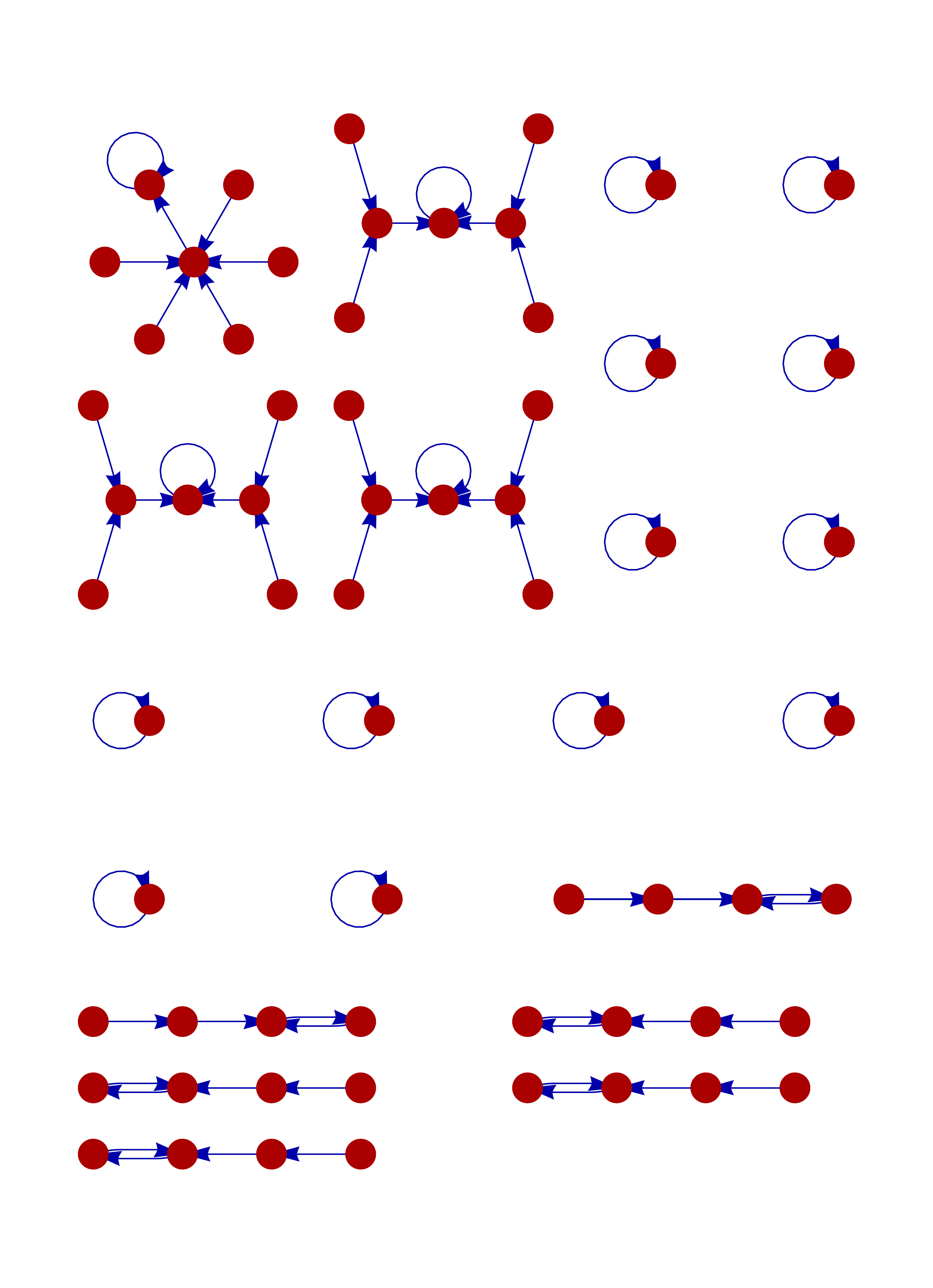}
        \caption{}
     \end{subfigure}
     \caption{The attractor and fixed point structure of (a) Rule 30 for 
                             $L=4$, (b) Rule 54 for $L=4$, (c) Rule 54 for $L=5$ and (d) Rule 201 for $L=6$. Red points represent the available states and the arrows the transitions between them, according to the given CA.}
     \label{fig:Rule_cycles_30_and_54_and_201}
 \end{figure*}

\begin{table}
    \fontsize{8}{11}\selectfont
    \begin{tabular}{SSSSSSS} \toprule
         \hspace{-0.24cm} $L$               & $M$  &&&  \\ \midrule
         3        &  {$1$}             \\
         4        &  {$8, 1          $} \\ 
         5        &  {$5, 1          $} \\
         6        &  {$1$}              \\
         7        &  {$63, 4, 1      $} \\
         8        &  {$40, 8, 1      $} \\
         9        &  {$171, 72, 1    $} \\
         10       &  {$15, 5, 1      $} \\
         11       &  {$154, 17, 1    $} \\
         12       &  {$102, 8, 3, 1  $}\\
         13       &  {$832, 260, 247, 91, 1 $} \\
         14       &  {$1428, 133, 112, 84, 63, 14, 4, 1 $} \\
         15       &  {$1455, 30, 9, 7, 5, 1   $}  \\
         16       &  {$6016, 4144, 40, 8, 1      $}  \\
         17       &  {$10846, 1632, 867, 306, 136, 17, 1 $} \\
         18       &  {$2844, 186, 171, 72, 24, 1   $}  \\
         19       &  {$3705, 247, 133, 38, 1       $}    \\
         20       &  {$6150, 3420, 1715, 580, 68, 30, 15, 8, 5, 1 $} \\
         21       &  {$2793, 597, 409, 63, 44, 42, 4, 1 $} \\
         22       &  {$3553, 3256, 781, 154, 77, 66, 17, 1 $}  \\
         23       &  {$38249, 4784, 138, 1 $} \\
         24       &  {$185040, 5448, 366, 312, 102, 40, 20, 8, 3, 1$} \\
         25       &  {$588425, 74525, 3470, 2950, 275, 5, 1$} \\  \bottomrule
    \end{tabular}
    \caption{\label{tab:periods_30}Periods for Rule 30.}
 \end{table}
 
 \begin{table}
    \fontsize{8}{11}\selectfont
    \begin{tabular}{SSSSSSS} \toprule
        \hspace{-0.24cm} $L$               &  $M$  &&&  \\ \midrule
         3        &  {$1$} \\
         4        &  {$4, 2, 1 $}  \\
         5        &  {$1      $}   \\
         6        &  {$4, 1   $}   \\
         7        &  {$4, 1   $}   \\
         8        &  {$8, 6, 4, 2, 1 $} \\
         9        &  {$27, 4, 1 $}  \\
         10       &  {$30, 4, 1 $}  \\
         11       &  {$99, 11, 4, 1 $} \\
         12       &  {$12, 10, 4, 2, 1 $} \\
         13       &  {$169, 4, 1 $}  \\
         14       &  {$112, 4, 1 $}  \\
         15       &  {$ 330, 4, 1 $}  \\
         16       &  {$40, 16, 14, 8, 6, 4, 2, 1 $} \\
         17       &  {$289, 51, 4, 1  $} \\
         18       &  {$306, 90, 180, 4, 1 $}  \\
         19       &  {$494, 437, 247, 57, 54, 27, 19, 4, 1 $} \\
         20       &  {$86, 60, 48, 32, 30, 24, 20, 18, 4, 2, 1 $} \\
         21       &  {$399, 147, 63, 14, 4, 1  $} \\
         22       &  {$484, 264, 242, 198, 121, 99, 32, 11, 4, 1 $} \\
         23       &  {$52371, 690, 575, 4, 1 $} \\
         24       &  {$312, 98, 56, 42, 32, 24, 22, 12, 10, 8, 6, 4, 2, 1 $} \\
         25       &  {$1800, 550, 115, 75, 4, 1$}  \\
         26       &  {$624, 546, 520, 338, 182, 169, 120, 78, 32, 26, 14, 4, 1 $} \\
         27       &  {$918, 837, 783, 459, 243, 81, 80, 27, 4, 1 $} \\
         28       &  {$224, 112, 110, 84, 64, 34, 32, 28, 26, 4, 2, 1$}  \\
         29       &  {$783, 725, 464, 203, 87, 80, 18, 4, 1 $}  \\
         30       &  {$780, 750, 660, 630, 420, 330, 150, 90, 75, 32, 30, 4, 1$}  \\
         31       &  {$1240, 1178, 1023, 961, 651, 450, 341, 217, 93, 42, 4, 1$}  \\
         32       &  {$608, 544, 144, 122, 96, 72, 40, 36, 32, 30, 22, 16, 14, 8, 6, 4, 2, 1$} \\
         33       &  {$1056, 1023, 957, 858, 429, 297, 143, 111, 99, 48, 11, 4, 1 $} \\
         34       &  {$952, 850, 816, 782, 578, 544, 289, 272, 119, 102, 51, 36, 34, 32, 4, 1$}  \\
         35       &  {$1540, 1470, 1225, 770, 735, 185, 105, 54, 35, 4, 1$} \\  \bottomrule
    \end{tabular}
    \caption{\label{tab:periods_54}Periods for Rule 54.}
 \end{table}

In Fig.~\ref{fig:Rule_cycles_30_and_54_and_201} we sketch the attractor structure for these nonlinear rules for some indicative system sizes, $N = L \times M$. Here $L$ refers to the linear size of the given CA. $M$ refers to the size in the time dimension that gives rise to nontrivial periodic orbits. More precisely, every red point corresponds to a configuration of the given system size shown. This means that, for example, for panel (a) there are $2^L = 2^4$ total configurations. Arrows indicate the one-step evolution of the given CA. Some states have self loops, indicating that they evolve to themselves under the given update rule (fixed points), while others form periodic structures of longer periods.

In Table \ref{tab:periods_30} we give the period lengths for Rule 30 up to size $L=25$ and  in Table~\ref{tab:periods_54} for Rule 54 up to $L=35$. The first column corresponds to the linear size of the CA, while the second column corresponds to the given number of time evolution steps which give rise to irreducible periodic structures for periodic boundaries in the time dimension. For example, Fig.~\ref{fig:rule30_tiles}a depicts one of the periodic orbits of Rule 30 with $M=8$. Since the CA rules are studied for periodic boundaries in both the space and time dimensions, further allowed periodic orbits with period length $M=8$ can be found by translating Fig.~\ref{fig:rule30_tiles}a in the time or space dimensions. Panel Fig.~\ref{fig:rule30_tiles}b and Fig.~\ref{fig:rule30_tiles}c show examples where the CA Rule 30 evolves into a trivial fixed point. This means that after one timestep the CA evolution returns to the same state (also exemplified in Fig.~\ref{fig:Rule_cycles_30_and_54_and_201}). However, in tables \ref{tab:periods_30} and \ref{tab:periods_54} we have not included the multiplicities of the periodic orbits for each give linear size $L$. We call these periodic orbits irreducible since we can equally create periodic orbits of, say, double the length of an orbit by replicating its evolution. The latter are excluded from our counting arguments.

In this way, we can infer the existence of a nontrivial number of ground states of the respective classical spin models. For example, starting from the classical energy function Eq.~\ref{non-onsite_rules_equations1}, we know that for PBC and for, say, a given system size $4\times 6$ the only ground states will come from the fixed points of the CA Rule 30. At the same time, for a system size $4 \times 16$ the ground states will originate in both the fixed points and the nontrivial period of Rule 30, since the system size is commensurate to the period length of Rule 30 for $L=4$.

Rule 201 has periods of length 2 and fixed points of length 1 of increasing number for all studied CA sizes. That's why we do not show its periodic structure explicitly. The number of these states with a given period length increase faster than linearly but always subextensively. For OBC the number of ground states depends on the extent of the constraint, being $2^{L + 2M - 2}$ for Rules 30, 54 and 201. For other rules where the constraint involves sites which belong in only two columns the degeneracy is $2^{L + M - 1}$.
 
The calculation of these limit cycles for the nonlinear rules cannot be performed in an algebraic-theoretic way as for the linear ones \cite{sfairopoulos2023boundary}, and one has to resort to brute-force enumeration. For this reason, we employ Floyd's tortoise and hare algorithm \cite{loehr2022the-tortise}. Another feature of the nonlinearity concerns the lack of predictive power on the cycle lengths. For additive CA and specific sequences of system sizes, there exist techniques for the prediction of their cycle lengths and respective theorems, see for example Refs.\cite{martin1984algebraic,1990_Ehrlich}. We do not expect any of these techniques to apply here. 

\balance
\bibliography{bibliography-17032025}

\end{document}